\documentclass[preprint]{elsarticle}

\usepackage{tabularx}
\usepackage{graphicx}
\usepackage{amsmath}
\usepackage{booktabs}
\usepackage{color}

\journal{ISPRS Journal of Photogrammetry and Remote Sensing}

\begin{document}


\begin{frontmatter}
%
\title{Potential of Nonlocally Filtered Pursuit Monostatic TanDEM-X Data for Coastline Detection}
%
%
%

\author{Michael~Schmitt$^1$, Gerald~Baier$^2$, Xiao~Xiang~Zhu$^{1,3}$}
        
\address{$^1$ Signal Processing in Earth Observation, Technical University of Munich (TUM), Germany\\
$^2$ Geoinformatics Unit, RIKEN Center for Advanced Intelligence Project, Tokyo, Japan\\
$^3$ Remote Sensing Technology Institute (IMF), German Aerospace Center (DLR), Wessling, Germany}        

\begin{abstract}
\textbf{\textit{This is the pre-print version, to read the final version please go to ISPRS Journal of Photogrammetry and Remote Sensing.}}\\
This article investigates the potential of nonlocally filtered pursuit monostatic TanDEM-X data for coastline detection in comparison to conventional TanDEM-X data, i.e. image pairs acquired in repeat-pass or bistatic mode. For this task, an unsupervised coastline detection procedure based on scale-space representations and $K$-medians clustering as well as morphological image post-processing is proposed. Since this procedure exploits a clear discriminability of ``dark'' and ``bright'' appearances of water and land surfaces, respectively, in both SAR amplitude and coherence imagery, TanDEM-X InSAR data acquired in pursuit monostatic mode is expected to provide a promising benefit. In addition, we investigate the benefit introduced by a utilization of a non-local InSAR filter for amplitude denoising and coherence estimation instead of a conventional box-car filter. Experiments carried out on real TanDEM-X pursuit monostatic data confirm our expectations and illustrate the advantage of the employed data configuration over conventional TanDEM-X products for automatic coastline detection. 
\end{abstract}


\end{frontmatter}


\section{Introduction}
From October 2014 to February 2015, the German synthetic aperture radar (SAR) satellite formation TanDEM-X \cite{Krieger2007} was operated in \textit{pursuit monostatic (PM) mode}, an innovative imaging mode allowing to acquire interferometric SAR data with increased spatial and temporal baselines \cite{Hajnsek2014,Lumsdon2015}. With an along-track distance of 76 km between both satellites (corresponding to a temporal baseline of about 10 s), the temporal decorrelation was expected to remain small for most terrain types -- except for water areas, which tend to decorrelate much faster due to the fast-changing wave appearance. Therefore, there is a great interest in a beneficial exploitation of pursuit monostatic InSAR data for water-related remote sensing applications, e.g. oil spill and ship detection \cite{Velotto2016}. In this context, the goal of this paper is to demonstrate the benefit of nonlocally filtered pursuit-monostatic TanDEM-X data for an improvement of the popular task of coastline detection.   

\subsection{Coastline Detection from SAR Imagery}\label{sec:CoastlineDetection}
The monitoring of coastal zones is an important task with implications to security, sustainable development, and environmental protection. Remote sensing has long been a powerful tool in that regard, especially in the context of coastline detection \cite{Gens2010}. Within this field, the exploitation of SAR data for coastline detection has met particular interest due to the favorable effect that water areas tend to appear dark, while land surfaces tend to appear bright in SAR amplitude images. Thus, pioneering work was already published almost three decades ago, when Lee et al. proposed a procedure employing speckle filtering, Sobel edge detection and edge tracing to extract coastlines from Seasat amplitude data \cite{Lee1990}. While subsequent works also focused on the exploitation of amplitude imagery and the corresponding need to cope with speckle noise \cite{Zhang1994,Descombes1996,Mason1996,Horritt1999,Niedermeier2000,Liu2004,Silveira2009}, the ERS-1/2 tandem configuration began to offer also coherence maps as potential input to coastline detection procedures \cite{Schwaebisch1997,Dellepiane2004}. Similar to the amplitude-based procedures, the underlying rationale for coherence-based approaches is the assumption that water surfaces will appear dark, while land surfaces will tend to appear bright. Based on the insights gained from these early investigations, later the authors began to exploit both amplitude and coherence imagery simultaneously \cite{Niedermeier2005,Wendleder2013}, finally giving way to the first approach using the original complex SAR data for statistically motivated coastline extraction \cite{Baselice2013}. 
It is noteworthy that recent publications still often resort to amplitude image analysis -- albeit now with modern data offering much higher spatial resolutions than decades ago \cite{Buono2014,Wiehle2015}. In addition, polarimetric information has become an additional source of information, which waits to be exploited beneficially \cite{Nunziata2014,Nunziata2016,Liu2017}. 

From a methodical point of view, two major groups can be identified: While the first group is based on traditional, bottom-up image analysis approaches including techniques such as speckle filtering, histogram thresholding, edge analysis, or Markov-random-field supported segmentation (e.g. \cite{Lee1990,Descombes1996,Schwaebisch1997,Dellepiane2004,Liu2004,Wendleder2013,Buono2014,Wiehle2015,Nunziata2016}), the second group resorts to level set segmentation in the form of active contours or snakes (e.g. \cite{Mason1996,Horritt1999,Niedermeier2000,Niedermeier2005,Silveira2009,Liu2017}). Exceptions to these major groups are approaches like the one proposed by Zhang et al. \cite{Zhang1994}, who use unsupervised clustering of first order statistical features or by Baselice et al. \cite{Baselice2013}, who propose a Gaussian Markov Random Field in order to estimate hyperparameters from complex multi-baseline SAR data, which are then used in a standard thresholding procedure for land-water segmentation. 

As an alternative to the existing approaches, which does not rely on coarse manual initialization of the desired contours, this paper employs a simple, unsupervised land-water segmentation strategy, which fuses both amplitude and coherence information achieved by nonlocal filtering \cite{Buades2005} of TanDEM-X PM data, using two-dimensional $K$-medians clustering \cite{Bradley1997} embedded in a scale space strategy \cite{Lindeberg1994}. As input to our approach, we intend to use nonlocally filtered InSAR data (i.e. amplitude and coherence images) because they provide both noise-free and still detailed information.

\subsection{Non-Local InSAR Filtering}
The concept of non-local means filtering was first presented to the Computer Vision community by Buades et al.\ in 2005~\cite{Buades2005}. The underlying rationale is that non-local filters exploit the high degree of redundancy in natural images: similar patches of every target patch often can be found many times in the same image. Thus, rather than averaging pixels in a pre-defined local neighborhood (e.g.\ within rectangular box-car windows), or in spatially connected adaptive regions, non-local filters consider pixels in a significantly larger search area and weight them according to a certain patch-based similarity measure~\cite{Deledalle2012}. Over the years, non-local filters have been shown to provide strong noise reduction capabilities while still well retaining fine structures such as linear features or point targets.

In 2011, the non-local paradigm was brought into the SAR interferometry community for the first time, when Deledalle et al.\ proposed the NL-InSAR algorithm, a procedure designed for non-local estimation of intensity, coherence and interferometric phase of an InSAR image pair~\cite{Deledalle2011}.
NL-InSAR inspired many remote sensing researchers to propose their own non-local InSAR denoising procedures (e.g.~\cite{Lin2015, Baier2016}), with NL-SAR a generalized version, which extends NL-InSAR from InSAR image pairs to stacks of multi-baseline or polarimetric SAR data~\cite{Deledalle2015} was proposed.
However, nonlocal filters do not come without drawbacks: the algorithms can produce filtering artifacts and are quite computationally expensive.
This paper aims to demonstrate the benefit of non-local InSAR filtering in the context of SAR-based coastline detection.

\subsection{Structure of this Paper}
The remainder of this article is structured as follows: Section~\ref{sec:TDXpm} will provide background information about the pursuit monostatic mode of the TanDEM-X mission. Section~\ref{sec:nlinsar} will summarize the employed non-local InSAR algorithm that is used to estimate denoised amplitude and coherence images from TanDEM-X PM data. After that, Section~\ref{sec:coastlineDetection} will describe the proposed unsupervised coastline detection procedure. In Section~\ref{sec:experiments}, experiments and results will be presented before the paper will be concluded by discussion and summary of the results in Sections~\ref{sec:discussion} and~\ref{sec:conclusion}. 


\section{TanDEM-X Pursuit Monostatic Mode}\label{sec:TDXpm}

\subsection{The Pursuit Monostatic Configuration}
\begin{figure}[htb]
\centering
\includegraphics[width=0.8\textwidth]{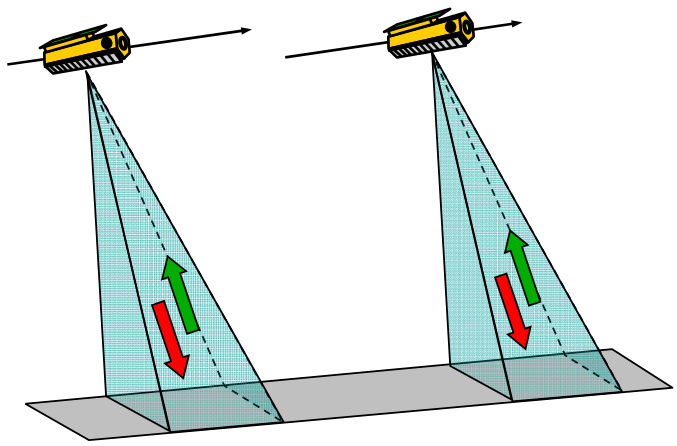}
\caption{Pursuit monostatic mode for TanDEM-X data acquisition \cite{Hajnsek2014}. Both the TerraSAR-X and the TanDEM-X satellites individually operate as monostatic emitters and receivers, while the distance between both satellites amounted to about 76 km. The resulting temporal baseline of 10 s is much longer than for bistatic acquisitions, but also significantly shorter than typical repeat-pass temporal baselines.}\label{fig:PM}
\end{figure}

Between October 2014 and February 2015, the TanDEM-X mission was steered to acquire interferometric SAR data in the so-called pursuit monostatic configuration, during which the TerraSAR-X and the TanDEM-X satellites were operated independently from each other (cf. Fig.~\ref{fig:PM}). It was characterized by a set of drifting across-track baselines spanning between 0 m and 750 m. Since these baselines were available at all latitudes and within short time periods, they were well suited for different scientific applications such as SAR tomography or large baseline investigations in the polar regions. The main feature of this configuration, however, was the comparably large along-track separation of 76 km (equalling a temporal baseline of 10 seconds) between the two satellites. Since X-band InSAR coherence of water surfaces quickly drops for temporal baselines larger than about 5 seconds \cite{Suchandt2017}, this ensured enhanced discriminability of water and land surfaces as required for the coastline detection pipeline proposed in this paper.

\subsection{Expected Benefit of Pursuit Monostatic Data in Land-Water Segmentation}\label{sec:CoherenceMatrix}

\begin{figure}[htb]
\centering
\includegraphics[width=0.8\textwidth]{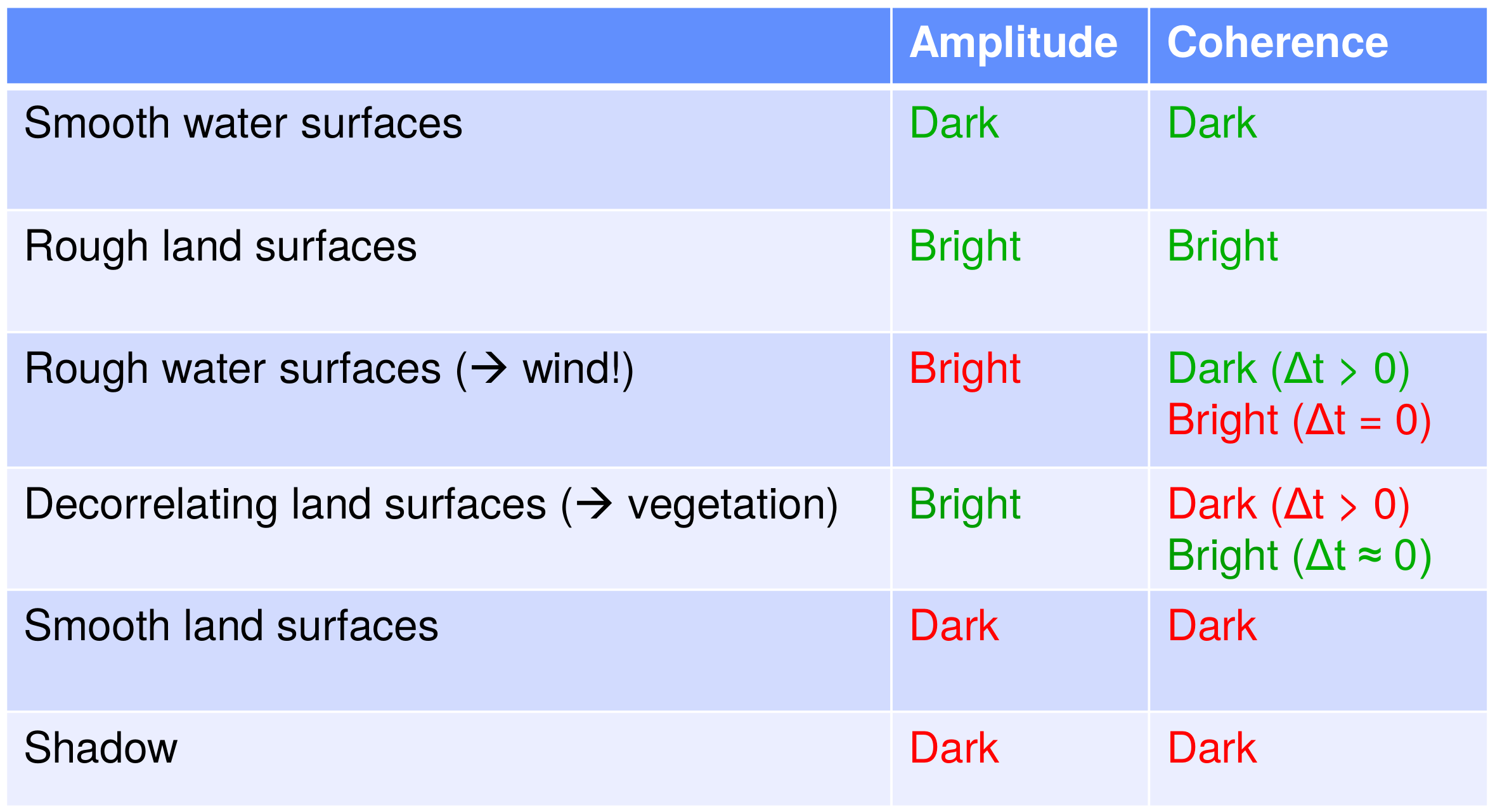}
\caption{Appearance of different surface types in amplitude and coherence images. Green entries mark favorable appearance supporting a clear distinction of land and water surfaces, while red entries mark unfavorable appearance that can lead to a confusion of land and water surfaces. It becomes obvious that the temporal baseline $\Delta t$ plays an important role for the usefulness of coherence as an indicator of land/water separation: If the baselines is non-zero but still very small, coherence data can greatly help to separate rough water surfaces from land or decorrelating vegetation from water.}\label{fig:CoherenceMatrix}
\end{figure}

Figure~\ref{fig:CoherenceMatrix} summarizes the appearance of different surface types in amplitude and coherence images. Since PM data provide a small non-zero temporal baseline $\Delta t \approx \varepsilon$, where $\varepsilon > 0$ is a small positive number, it improves the capability to distinguish between wind-affected (i.e. rough) water surfaces and land surfaces. In addition, it helps to avoid the confusion caused by decorrelating land surfaces (that would also appear dark in classical repeat-pass data) and smooth water surfaces. Figure~\ref{fig:CoherenceMatrix} also shows that a joint exploitation of amplitude and coherence images will further support the distinction capabilities. The only remaining problem cases are smooth land surfaces (e.g. paved roads) and areas affected by radar shadowing. Similar to smooth water surfaces, these surface types also act as mirrors to the radar signals and provide no backscattering power, thus leading to dark appearance. This, however, is a SAR-inherent effect that is independent of the specifically used imaging mode.


\section{Nonlocal InSAR Coherence Estimation}\label{sec:nlinsar}

\begin{figure}[htb]
\centering
\includegraphics[width=0.4\columnwidth]{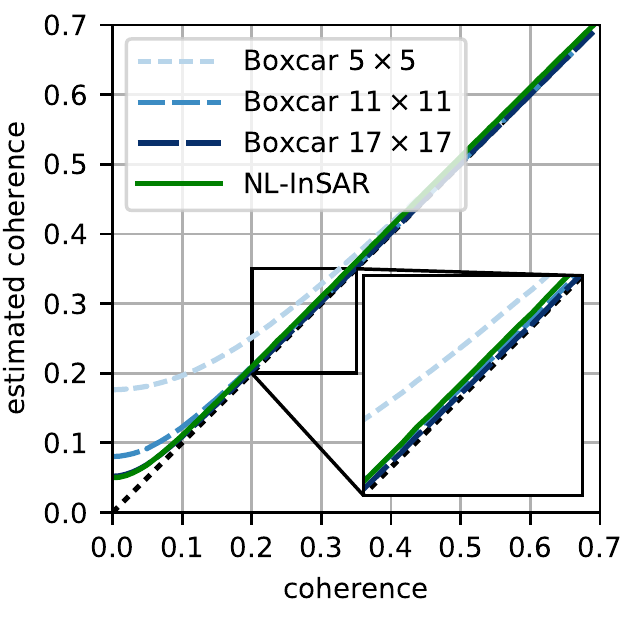}
\caption{True and estimated coherence with different estimators. The high number of looks of the nonlocal filter lead to a mostly unbiased estimate.}\label{fig:coh_est_bias}
\end{figure}

The appeal of nonlocal filters for our application are their less biased coherence estimates~\cite{Zhu2014}.
For mostly incoherent areas, such as water, estimators with a low number of looks overestimate the coherence~\cite{Touzi1999}.
Figure~\ref{fig:coh_est_bias} depicts the true and estimated coherence for a simulation with different estimators.
Clearly for low coherent areas a large estimator footprint is required.
In particular, significantly less biased coherence estimates can be achieved with the employed nonlocal filter, which serves as a very important input for coastline detection, as highlighted in Section~\ref{sec:CoherenceMatrix}.

Nonlocal filters also provide the additional benefit that the delineation of coastlines is not smoothed out as with conventional estimators with a fixed footprint.

\begin{figure}[htb]
\centering
\newcommand{\subfigwidth}{0.28\columnwidth}
\begin{tabular}{ccc}
\includegraphics[width=\subfigwidth]{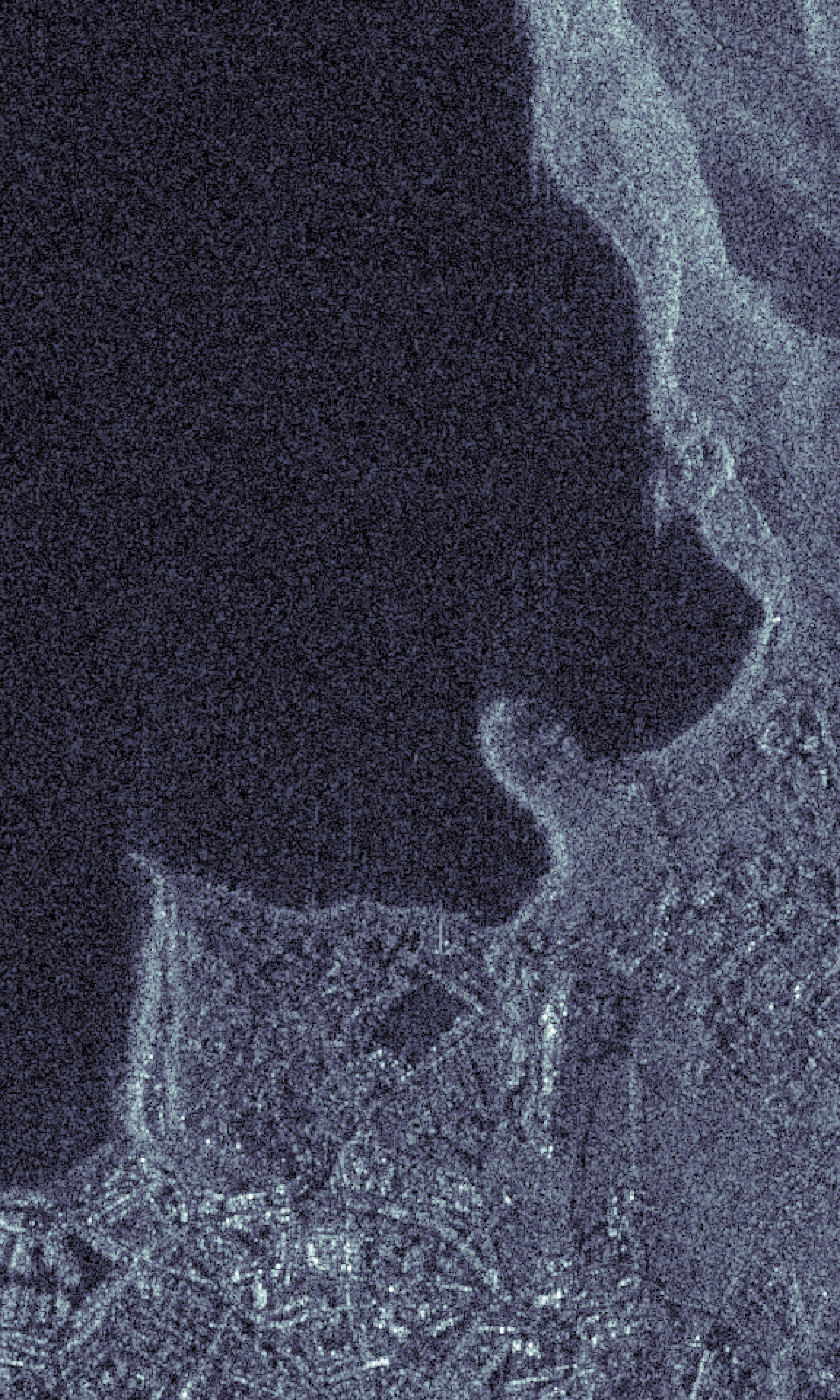} & \includegraphics[width=\subfigwidth]{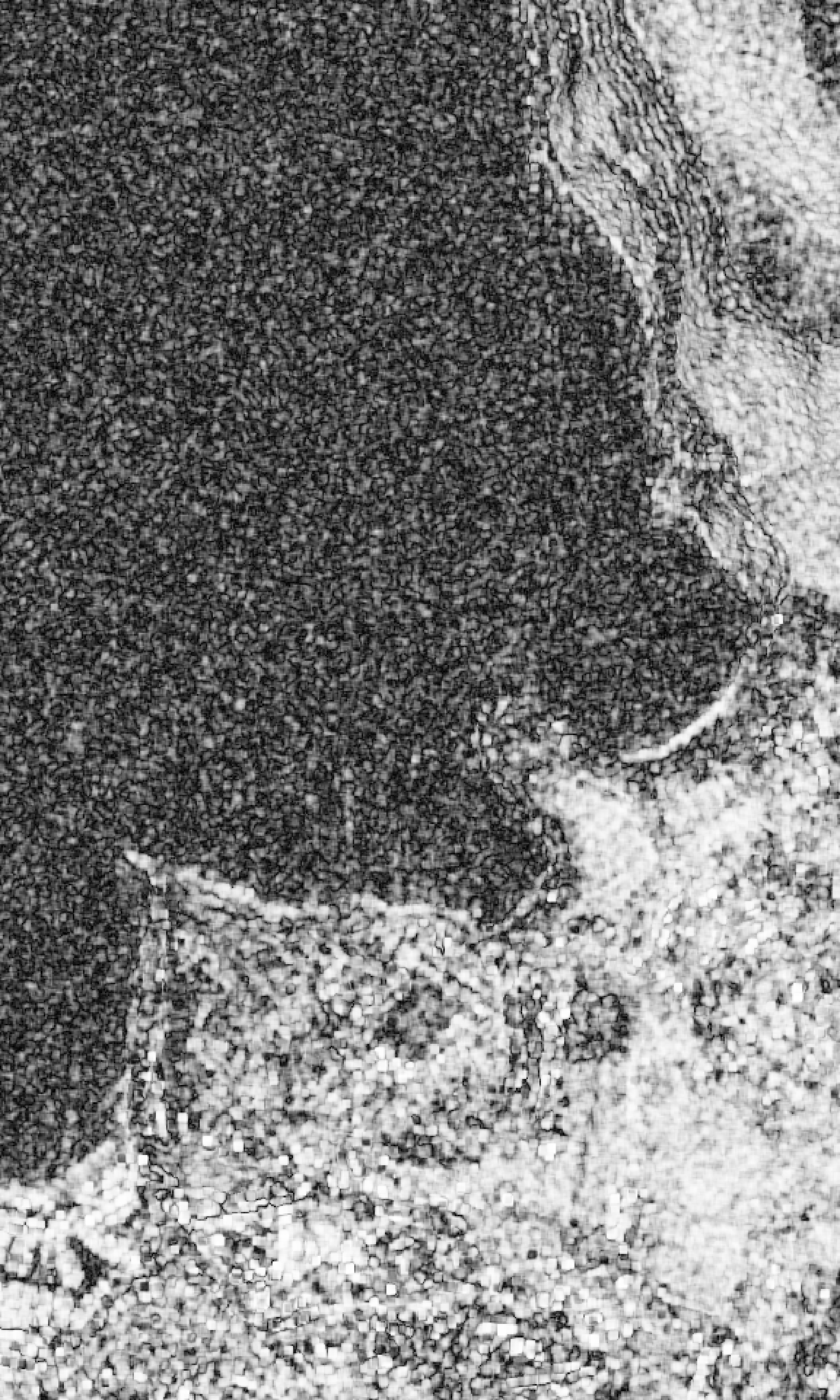} & \includegraphics[width=\subfigwidth]{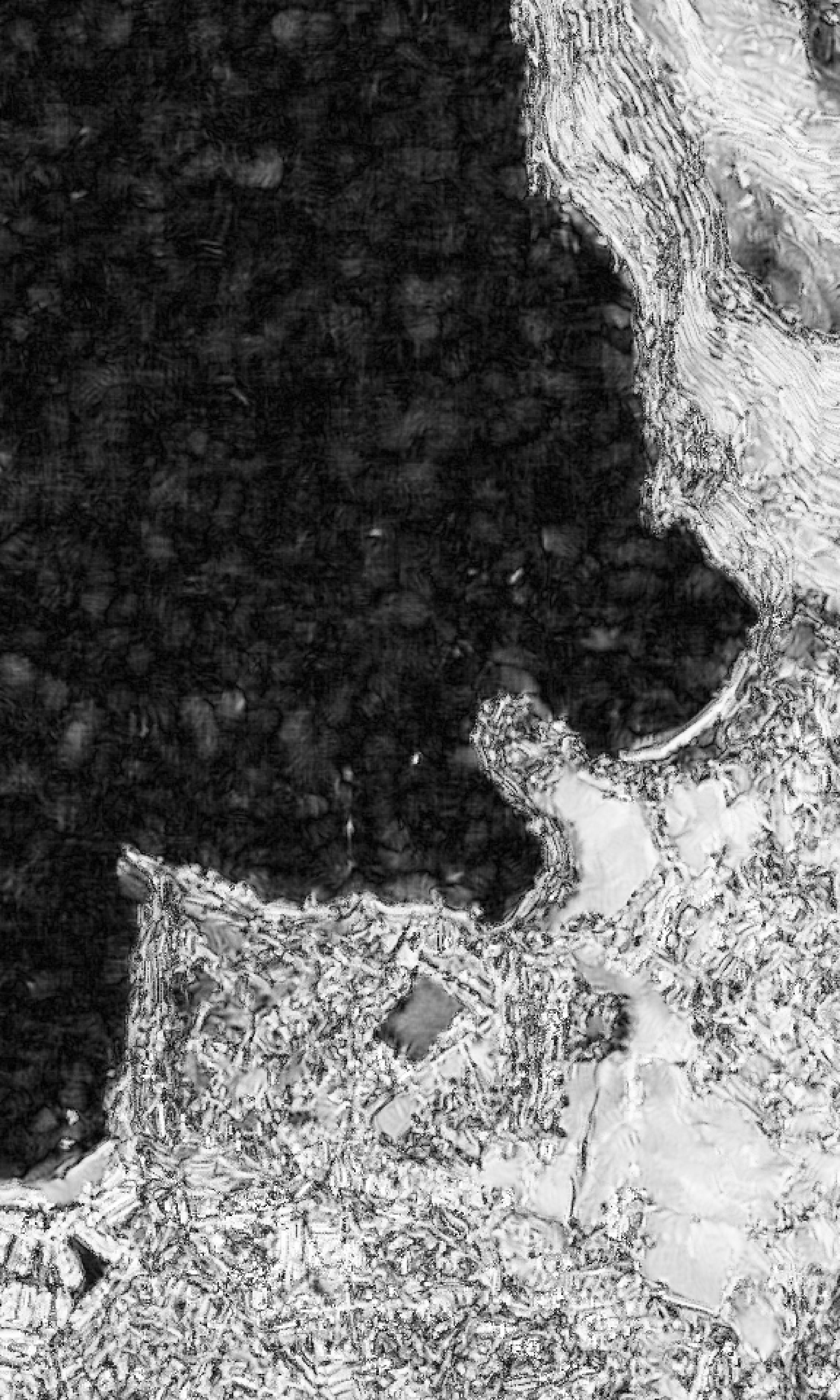} \\
(a) & (b) & (c)
\end{tabular}
\caption{comparison of coherence estimates (a) amplitude, (b) boxcar estimate, and (c) nonlocal estimate. The higher number of looks leads to a less biased coherence estimate}\label{fig:coh_est_comparison}
\end{figure}

Figure~\ref{fig:coh_est_comparison} shows for a coastal region the gain that can be achieved by employing nonlocal filters.
The boxcar estimate was computed on a $5 \times 5$ window, showing far more clutter and on average higher values for water than the nonlocally computed estimate, which relied on NL-InSAR~\cite{Deledalle2011}.

We thus also analyze the impact that nonlocal filtering as a preprocessing step has on the accuracy of the coastline detection.
In our analysis we employ NL-InSAR, yet similar conclusions could be drawn for other nonlocal filters, for example~\cite{Deledalle2015, Baier2016}.
The search window size and patch size has been set to 21 by 21 and 7 by 7, respectively.

\section{Unsupervised Coastline Detection}\label{sec:coastlineDetection}
\begin{figure*}[tbh]
\centering
\includegraphics[width=\textwidth]{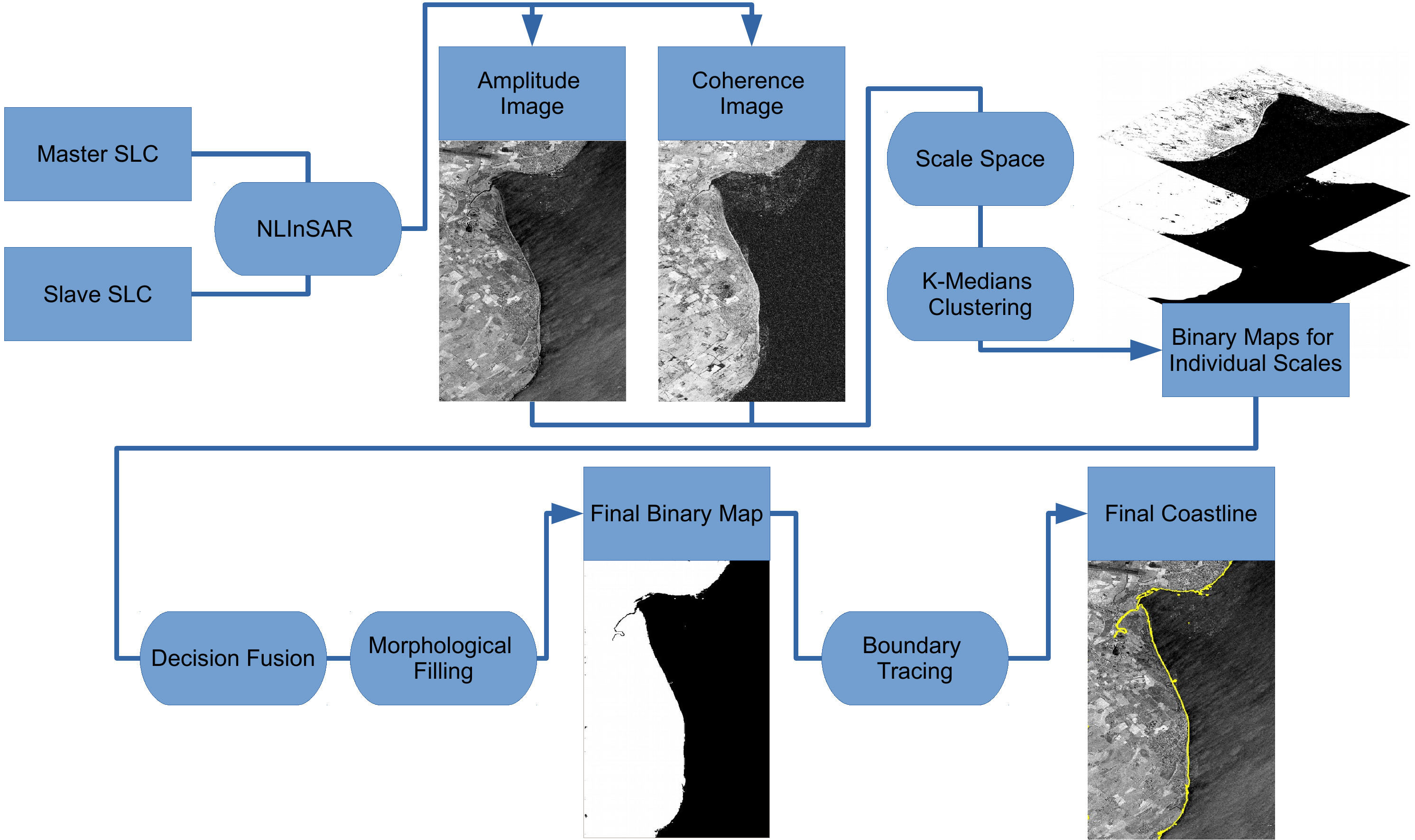}
\caption{Flowchart of the unsupervised coastline detection procedure presented in this paper.}\label{fig:flowchart}
\end{figure*}
Since the goal of this paper is to demonstrate the benefit of TanDEM-X data acquired in pursuit monostatic mode over TanDEM-X data acquired in the conventional bistatic or repeat-pass modes, we propose a simple coastline detection procedure that exploits the assumption of \textit{dark}-appearing water surfaces and \textit{bright}-appearing land surfaces as described in Section~\ref{sec:CoherenceMatrix}, which will be used as experimental framework. The flowchart of the procedure is depicted in Fig.~\ref{fig:flowchart}. Its core consists of a scale space representation of both the amplitude and the coherence image of the TanDEM-X scene, a $K$-Medians clustering step on each scale, where amplitude and coherence information are used as a 2D feature representation allowing the distinction of pixels appearing \textit{dark} in both amplitude and coherence (i.e. corresponding to the \textit{water} class) and pixels appearing \textit{bright} in both amplitude and coherence (i.e. corresponding to the \textit{land} class). Finally, the scale-space results are reduced to a single binary map. In classical data fusion terminology, this can be described as a combination of \textit{signal-level fusion} (i.e. the fusion of amplitude and coherence observations within each individual scale) and \textit{decision-level fusion} (i.e. the fusion of the intermediate results achieved on each scale to a final result) \cite{Schmitt2016}. 

The binary map achieved by this two-step data fusion procedure is then post-processed using morphological operators to produce the final land-water map from which subsequently the coastlines can be extracted. The individual steps of the procedure are described in detail in the following sections. 

\subsection{Scale Space}\label{sec:ScaleSpace}
The first step of the proposed coastline detection procedure is to calculate a scale space representation for both the amplitude and the coherence images of the filtered TanDEM-X pursuit monostatic dataset. 

\begin{figure}[htb]
\centering
\includegraphics[width=0.4\textwidth]{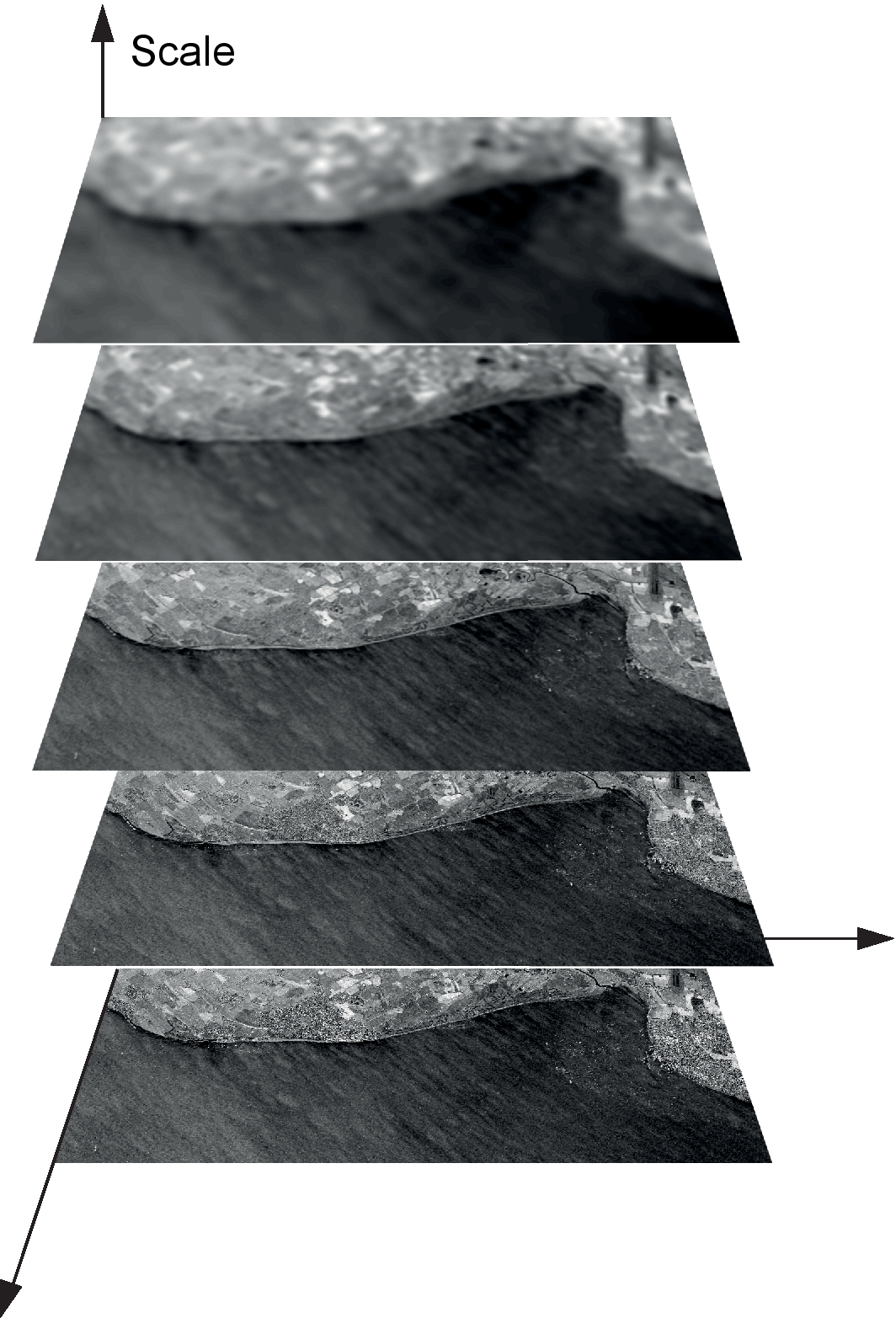}
\caption{Exemplary illustration of the scale space representation of a SAR amplitude image with five scales. The size of the Gaussian kernel increases logarithmically from bottom to top.}\label{fig:ScaleSpace}
\end{figure}

In short, the basic principle of scale space is to create new versions of the input image at different scales, where details decrease the more coarse the scale gets \cite{Lindeberg1994}. The difference to a pyramid image representation is that the scale space representation maintains the same image size at all scales (cf. Fig.~\ref{fig:ScaleSpace}), while pyramid representations lead to reduced image sizes at each pyramid level. Scale space is a widely employed concept in image processing and computer vision, which found its potentially most prominent application as part of the \textit{SIFT} algorithm \cite{Lowe2004}, where scale space representations of an input image are the basis for subsequent key point detection.

Mathematically, the scale space representation of an image can be achieved by ``blurring'' the image $I(x,y)$ via convolution with a Gaussian kernel $G(x,y,\sigma) = \frac{1}{2\pi\sigma^2}e^{-\left(x^2+y^2\right)/2\sigma^2}$, i.e.
\begin{equation}\label{eq:ScaleSpace}
L(x,y,\sigma) = G(x,y,\sigma)\ast I(x,y), 
\end{equation}
where (\ref{eq:ScaleSpace}) is applied several times with increasing variance $\sigma ^2$ of the Gaussian kernel to achieve a full scale space representation as shown in Fig.~\ref{fig:ScaleSpace}. For the work presented in this paper, we create the scale space with logarithmically increasing variance, i.e. $\sigma^2 \in \{0, 1, 4, 16, 64, 256, 1024, 4096\}$.  

The reason we include the scale space concept into the coastline detection pipeline proposed in this paper is that it enables robustness by increasingly filtering out noise components at higher scales while preserving fine details at lower scales. 
  
\subsection{$K$-Medians Clustering}
\begin{figure}[htb]
\centering
\includegraphics[width=0.8\textwidth]{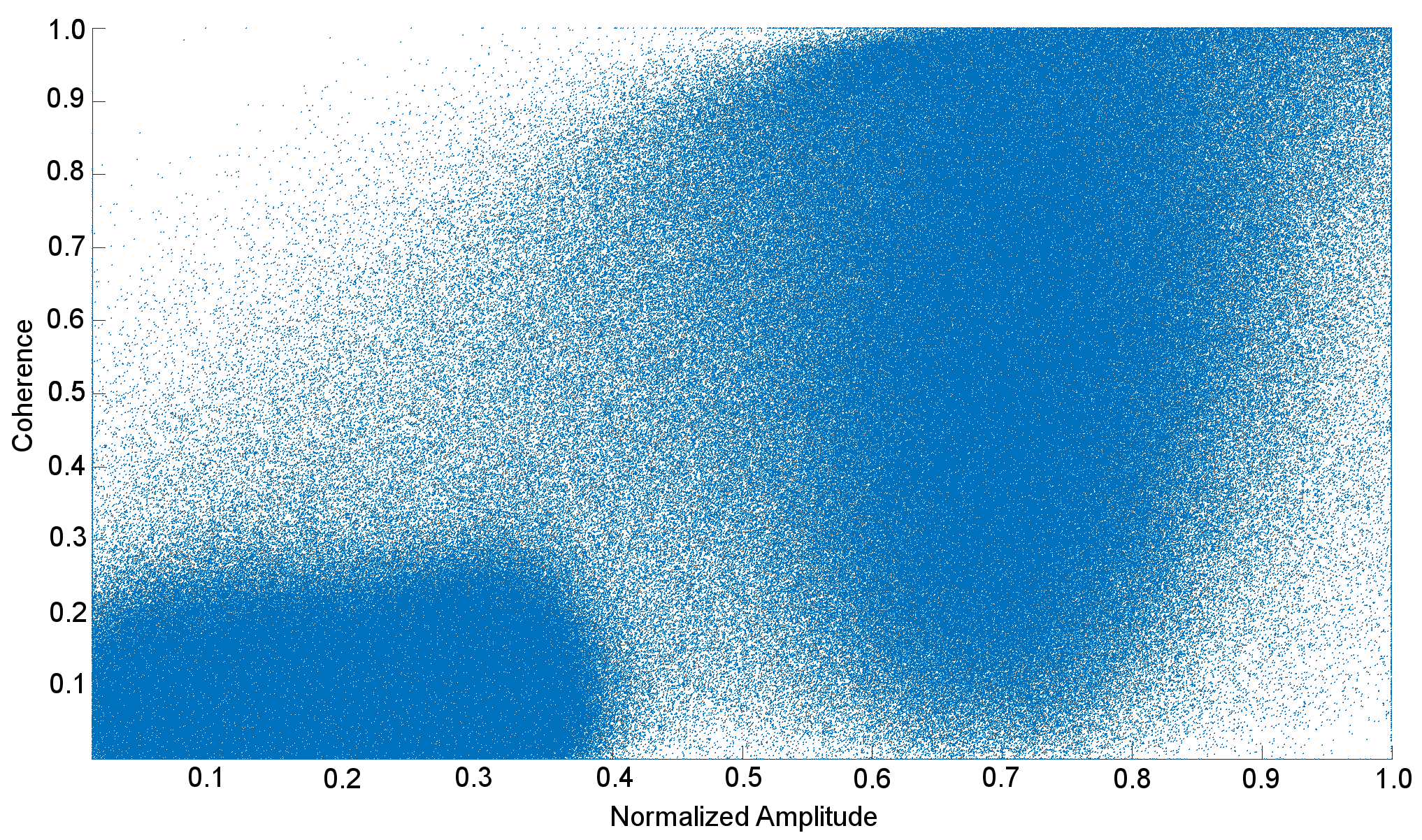}
\caption{Exemplary visualization of the two-dimensional distribution of normalized amplitude and coherence data for a scene containing approximately equal amounts of land and water surfaces. The two clusters for \textit{water} pixels (lower left) and \textit{land} pixels (upper right) are clearly visible.}\label{fig:ClusterAnalysis}
\end{figure}

After the scale space for both the amplitude and the coherence image have been created, we employ a $K$-medians clustering step at every scale, where each pixel is considered a data point described by two variables, namely its amplitude and its coherence value. As described in Section~\ref{sec:TDXpm} and confirmed by Fig.~\ref{fig:ClusterAnalysis}, this makes sense because \textit{water} pixels tend to appear \textit{dark}, whereas \textit{land} pixels tend to appear \textit{bright} in both amplitude and coherence. Fusing both kinds of information in this binary clustering step is supposed to improve the robustness to brightness changes not related to class distinction, because of the considerations depicted in Fig.~\ref{fig:CoherenceMatrix}: In the difficult cases (rough water surfaces or decorrelating vegetation), looking at amplitude only will lead to wrong clustering results, while adding coherence information acquired with small temporal baselines as provided by pursuit-monostatic data helps to ensure a correct clustering.

$K$-medians can be described as a variation of the better-known $K$-means clustering \cite{Jain2010}, where the median instead of the mean is used for calculating the cluster centroids. In essence, this leads to an $L_1$-norm minimization of the clustering error over all clusters instead of the $L_2$-norm. We use $K$-medians instead of $K$-means in order to enhance the robustness of this clustering step in the sense of robustness against potential outliers.

\subsection{Decision Fusion}\label{sec:DecisionFusion}
After carrying out the $K$-medians clustering for every scale of the amplitude/coherence dataset, $N$ (where $N$ is the number of scales in the scale space, here $N = 8$) binary maps exist, which can be seen as intermediate land/water classification results. These results are fused in a final \textit{decision fusion} step, which is implemented by the simple rule
\begin{equation}\label{eq:DecisionFusion}
c = 
\begin{cases}
      {land}, & \text{if}\ \sum_{i=1}^N b_i \geq 0.75\cdot N \\
      {water}, & \text{otherwise}
    \end{cases}
\end{equation}

Here, $b_i$ describes the binary clustering results, where $b_i = 1$ refers to the land cluster, and $b_i = 0$ refers to the water cluster. This means that the decision fusion step considers every pixel as a \textit{land} pixel that was found to be part of the \textit{land} cluster in at least 75\% of the scales. All other pixels are considered to be part of the \textit{water} class. The choice of the 75\%-threshold is based on the consideration that we wanted to create a stricter version of classical majority voting approaches. Thus, instead of accepting the result that collected most votes -- which would correspond to a threshold of 50\% -- we demand the result to be more distinct. 

\subsection{Post-Processing}
The binary land/water map resulting from the procedure described in Sections~\ref{sec:ScaleSpace} to \ref{sec:DecisionFusion} still contains some noisy pixels as well as ``holes'' in the land areas, which are, e.g., caused by inland water bodies. In order to remove both noise and inland water bodies, post-processing by a morphological fillhole algorithm \cite{Soille2004} is carried out. In short, the algorithm removes all minima, which are not connected to the image border. The result is a clean land/water map, from which the coastlines can be extracted by boundary tracing, e.g. using the Moore-Neighbor tracing algorithm modified by Jacob's stopping criteria, as it is implemented in MATLAB's Image Processing Toolbox \cite{Gonzalez2004}.

\section{Experiments \& Results}\label{sec:experiments}

\subsection{Test Data}
For the experiments in this paper, we make use of TanDEM-X pursuit monostatic image pairs acquired for two distinctly different scenes. The first dataset shows a part of the English Channel (EC), the second one shows the skerry coast next to Stockholm (ST). While the EC dataset is characterized by a rather simple coastline and a clear half/half distribution of relatively homogeneous land and water bodies, the ST dataset is characterized by an abundance of small islands and rather complex coastlines. Therefore, we are convinced that our results will provide a good indication about the general applicability of our findings when transferred to other coastline areas around the globe.

\begin{figure}[htb]
\centering
\begin{tabular}{cc}
\includegraphics[width=0.3\textwidth]{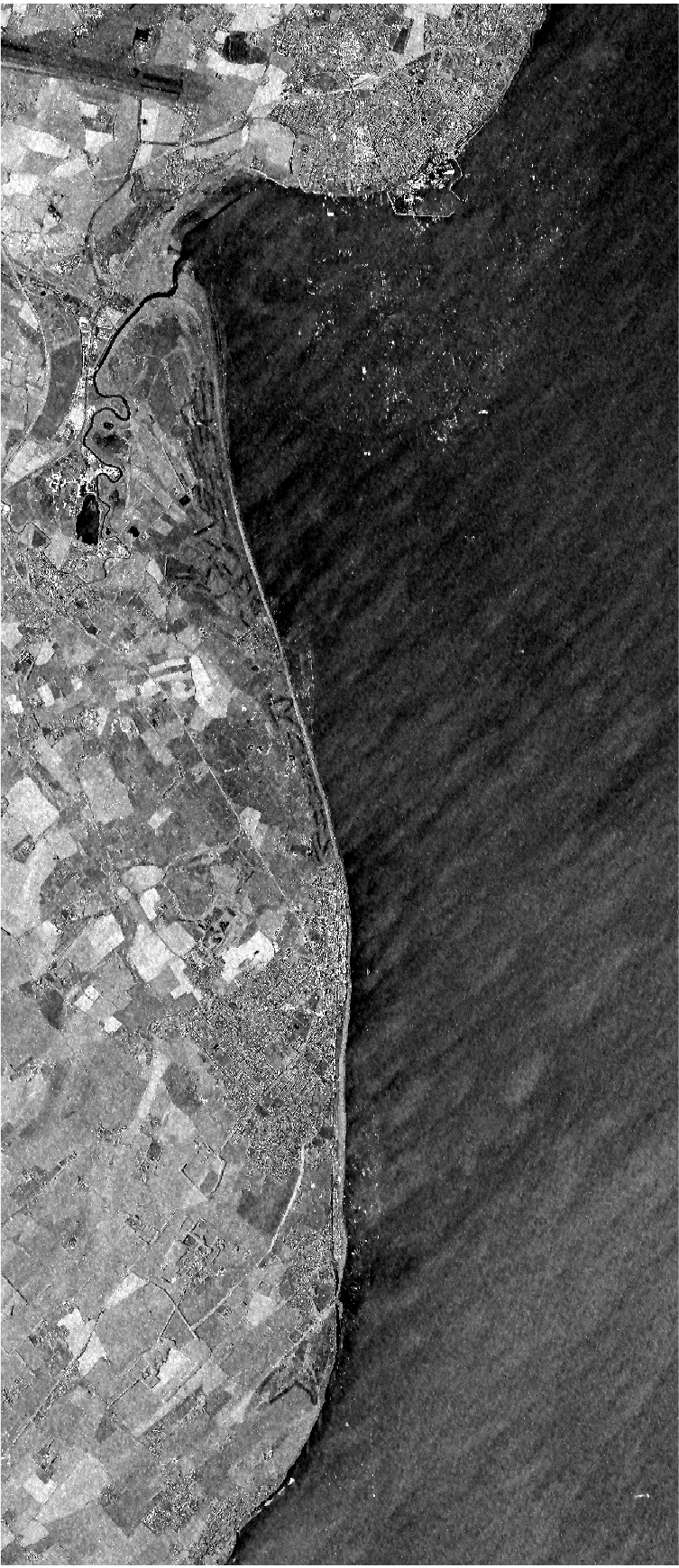} & \includegraphics[width=0.3\textwidth]{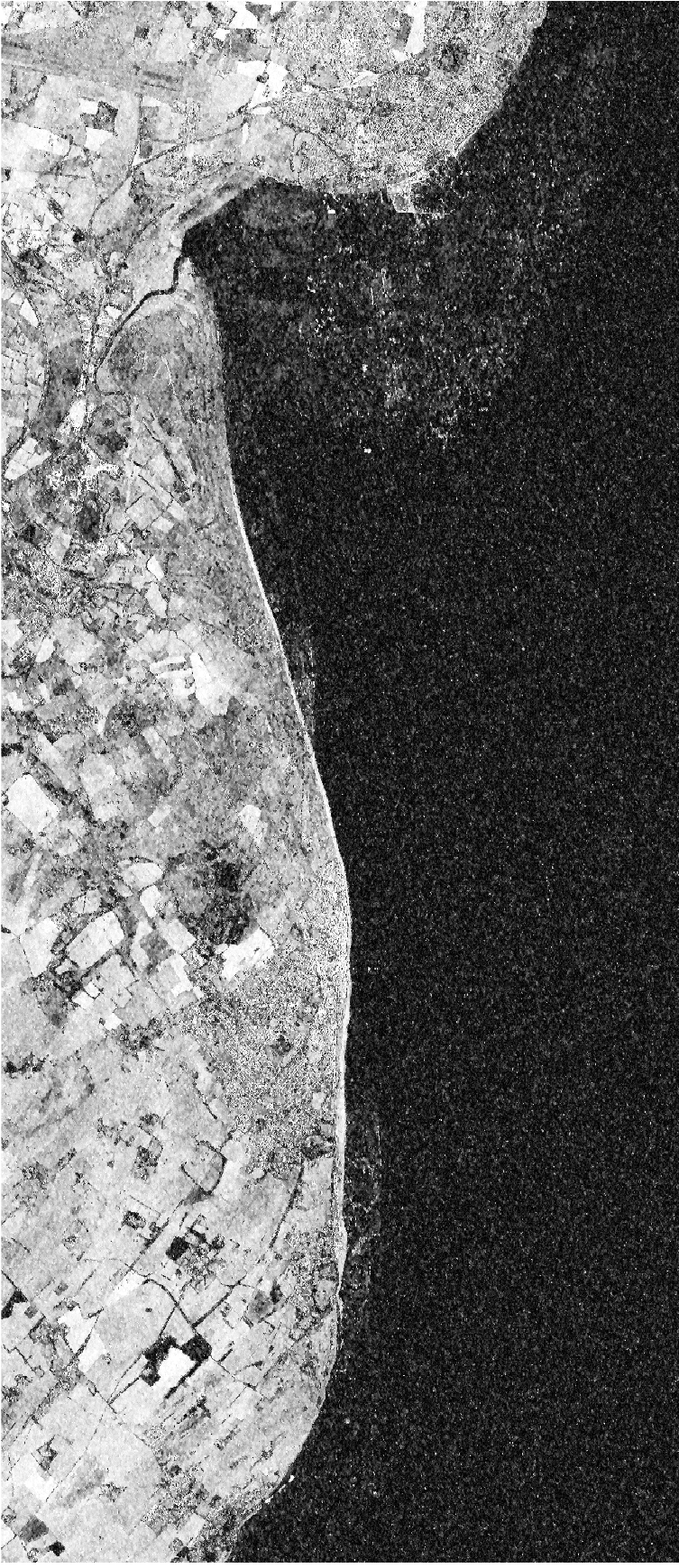}\\
(a) & (b)
\end{tabular}
\caption{Test subset English Channel (EC): Nonlocally filtered (a) amplitude and (b) coherence images acquired in pursuit monostatic mode. Note the azimuth ambiguities \cite{Li1983} in the upper right part of the scene.}\label{fig:EC_input}
\end{figure}

\begin{figure}[htb]
\centering
\begin{tabular}{cc}
\includegraphics[width=0.3\textwidth]{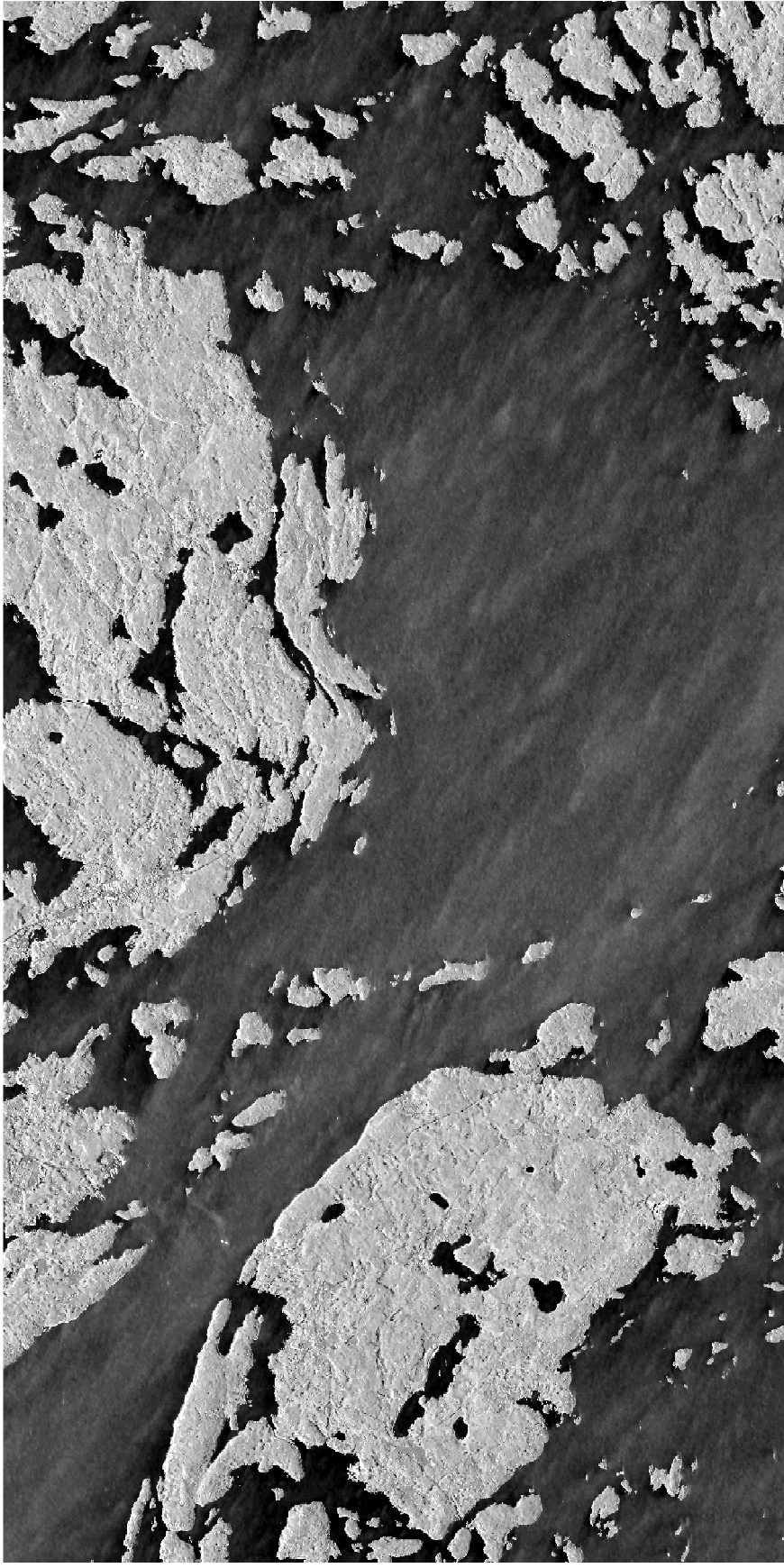} & \includegraphics[width=0.3\textwidth]{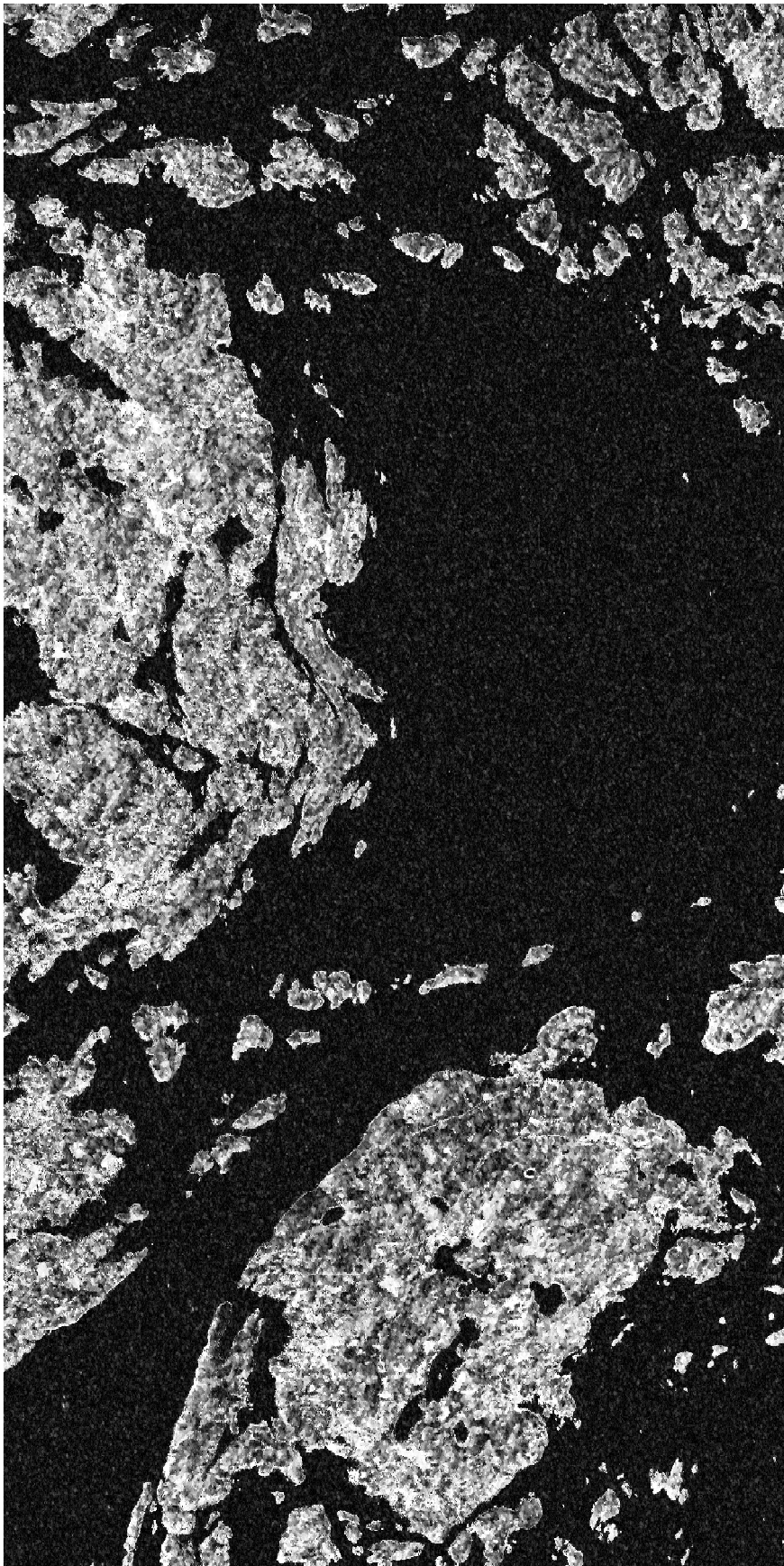}\\
(a) & (b)
\end{tabular}
\caption{Test subset Stockholm skerry coast (ST): Nonlocally filtered (a) amplitude and (b) coherence images acquired in pursuit monostatic mode.}\label{fig:ST_input}
\end{figure}

Details about the individual acquisitions we used can be found in Tab. \ref{tab:Data}. Both datasets were acquired in stripmap mode from an ascending orbit. After SAR raw data focusing, both the amplitude and coherence images extracted by the nonlocal filter described in Section~\ref{sec:nlinsar} have been geocoded using DLR's IWAP processor and then transformed and resampled to UTM coordinates with a square pixel spacing of 1.99 m (EC) and 1.61 m (ST), respectively. Finally, suitable subsets were extracted for the presented investigations. For both test subsets, the input images (i.e. nonlocally filtered amplitude and coherence images) are depicted in Figs.~\ref{fig:EC_input} and \ref{fig:ST_input}, respectively.

\begin{table*}[tbh]
\centering
 \footnotesize
  \caption{Acquisition Parameters of TanDEM-X Test Data}\label{tab:Data}
\begin{tabular}{lcccccc}
\toprule
&\multicolumn{2}{c}{\textit{Resolution (m)}}&\textit{Pixel Spacing (m)}&\textit{Incidence Angle}  &\multicolumn{2}{c}{\textit{Acquisition Time (UTC)}}\\
&\textit{Azimuth} & \textit{Range} &\textit{after geocoding}& \textit{at Scene Center ($^\circ$)}&  \textit{Master} & \textit{Slave}\\
\textbf{EC} & 6.60 & 1.77 & 1.99 & 44.95 & 2014-12-14T17:44:14 & 2014-12-14T17:44:24\\
\textbf{ST} & 3.30 & 1.18 & 1.61 & 37.18 & 2014-10-14T16:20:16 & 2014-10-14T16:20:26\\
    \bottomrule
\multicolumn{7}{l}{\scriptsize{EC: English Channel, ST: Stockholm}}    
\end{tabular}
\end{table*}

\subsection{Qualitative Results}\label{sec:qualResulPM}
\begin{figure}[htb]
\centering
\begin{tabular}{ccc}
\includegraphics[width = 0.3\textwidth]{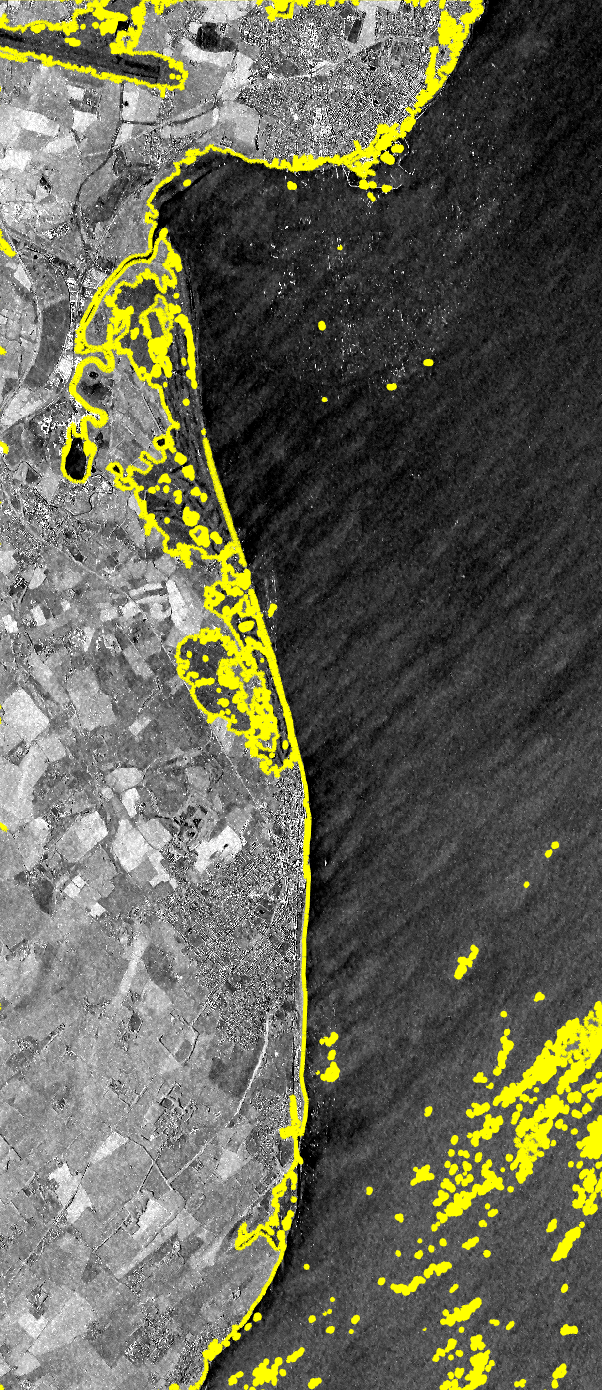} & \includegraphics[width = 0.3\textwidth]{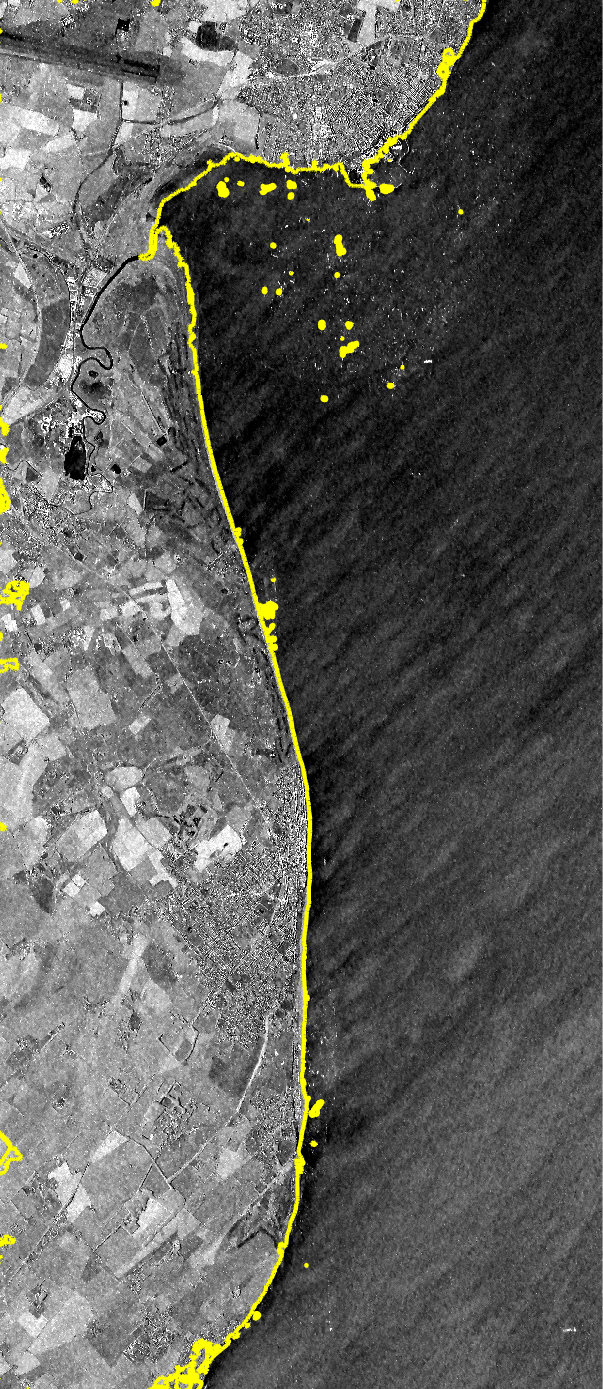} & \includegraphics[width = 0.3\textwidth]{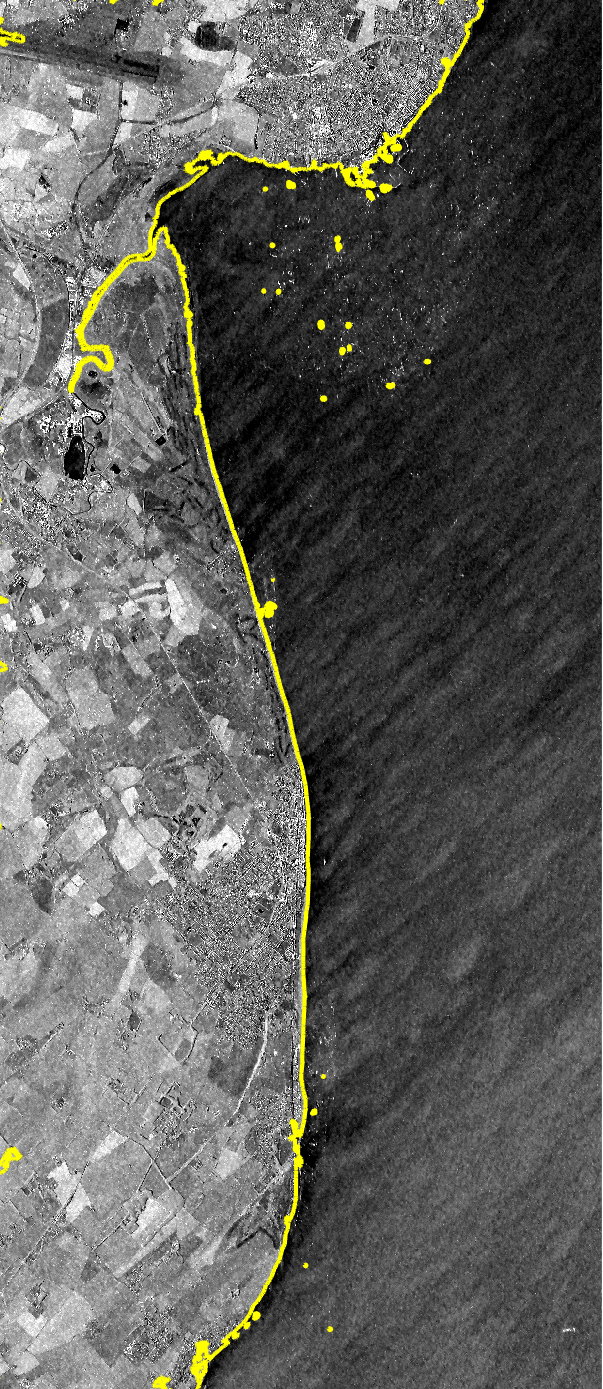}\\
(a) & (b) & (c)
\end{tabular}
\caption{Coastlines extracted from (a) the amplitude image, (b) the coherence image, and (c) from a fusion of amplitude and coherence of the EC dataset. The coastlines are overlayed to the amplitude image in all three cases.}\label{fig:EC_PM_nlinsar_results}
\end{figure}

\begin{figure}[htb]
\centering
\begin{tabular}{ccc}
\includegraphics[width = 0.3\textwidth]{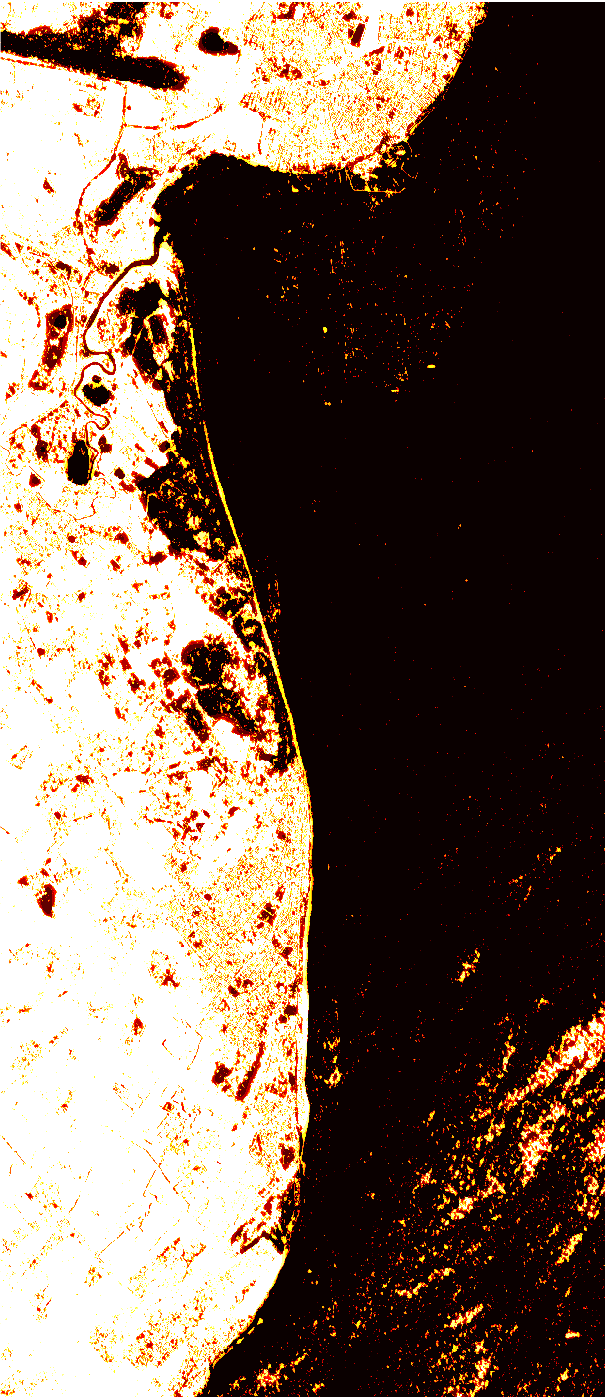} & \includegraphics[width = 0.3\textwidth]{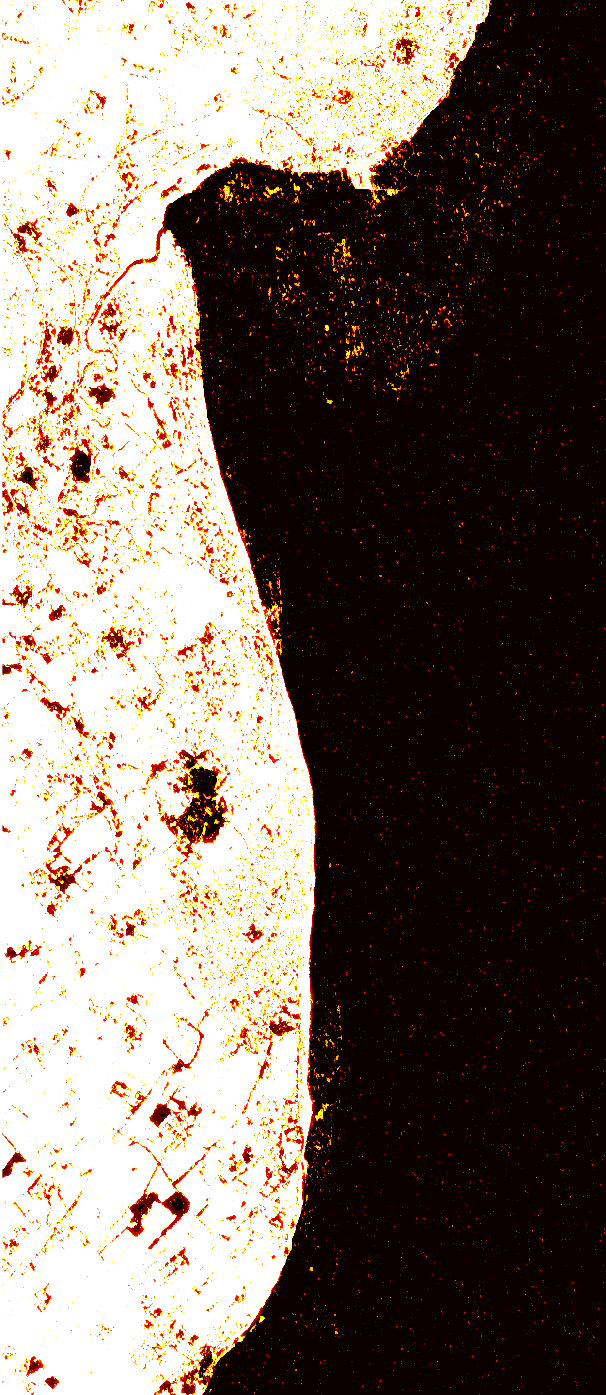} & \includegraphics[width = 0.3\textwidth]{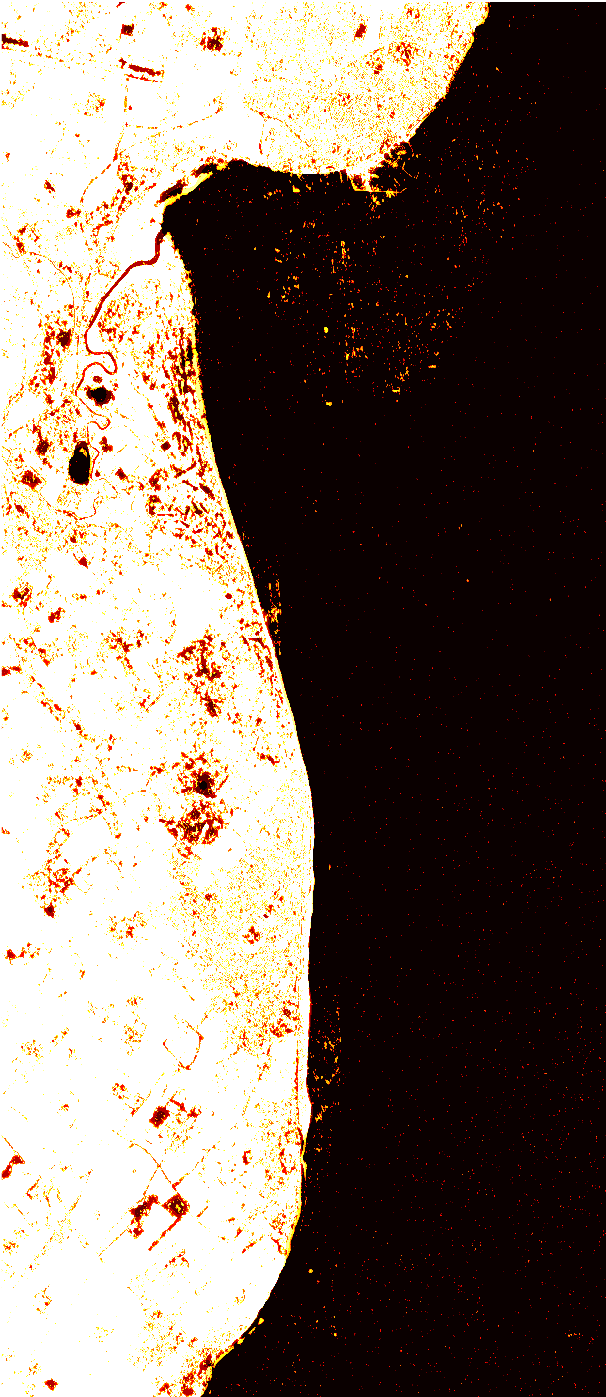}\\
(a) & (b) & (c)\\
\multicolumn{3}{c}{\includegraphics[width=0.9\textwidth]{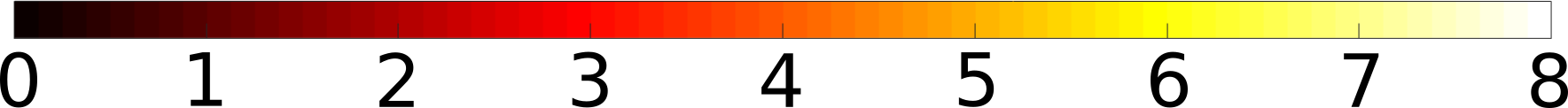}}
\end{tabular}
\caption{Sum of the binary segmentation results corresponding to the individual scales of the scale-space representation. The maps correspond to (a) the amplitude image, (b) the coherence image, and (c) the fusion of amplitude and coherence of the EC dataset.}\label{fig:EC_PM_Heatmaps}
\end{figure}

\begin{figure}[htb]
\centering
\begin{tabular}{ccc}
\includegraphics[width = 0.3\textwidth]{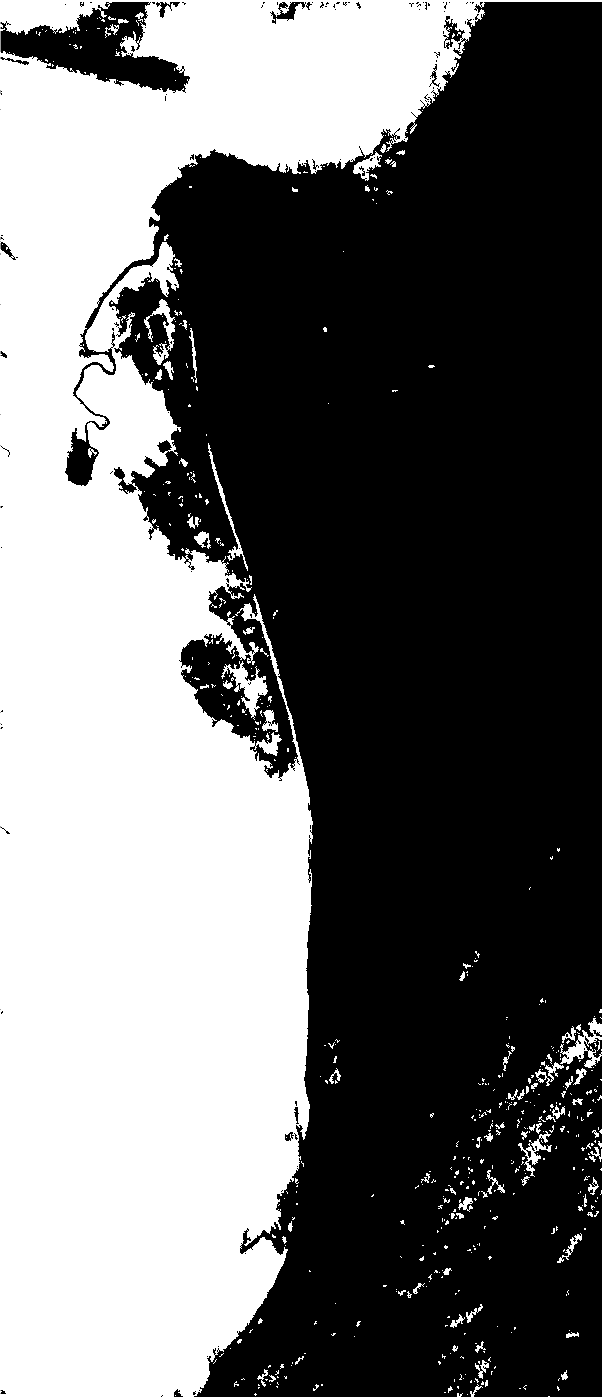} & \includegraphics[width = 0.3\textwidth]{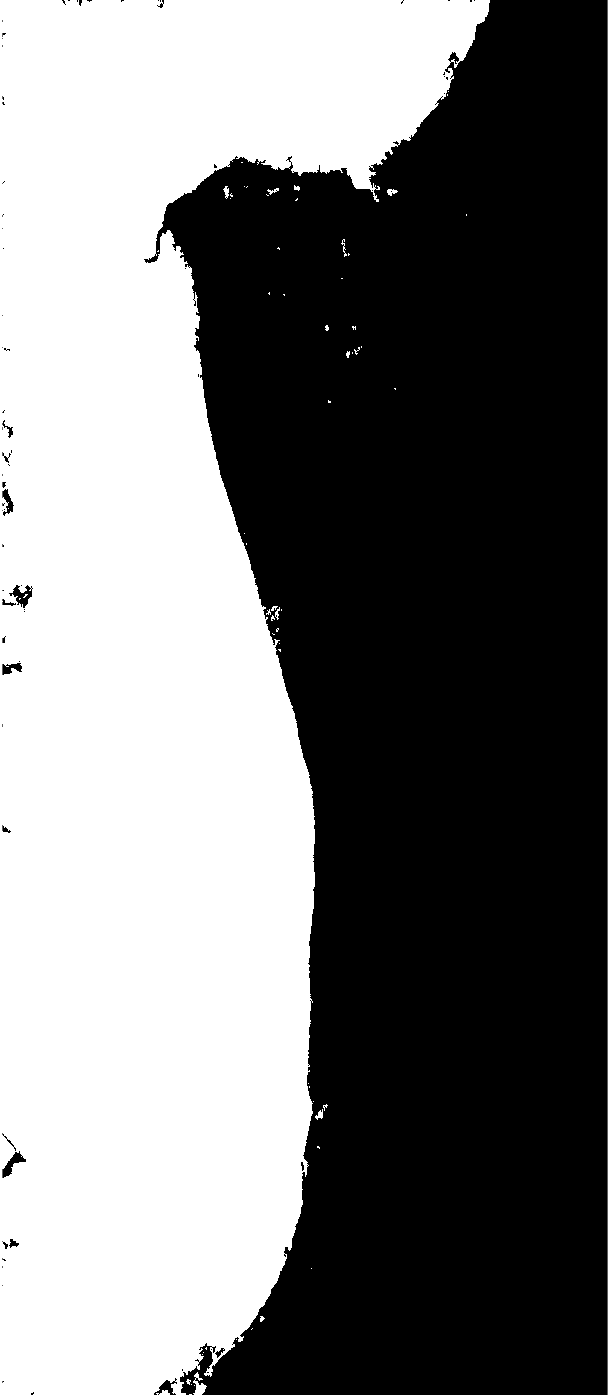} & \includegraphics[width = 0.3\textwidth]{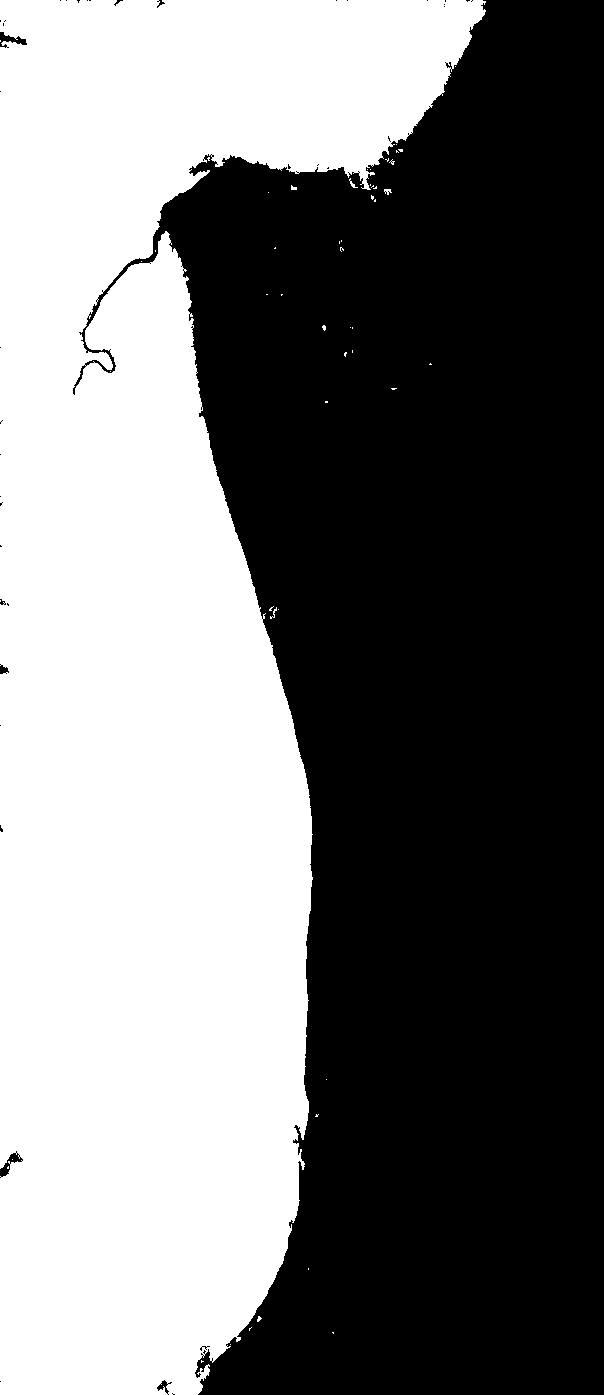}\\
(a) & (b) & (c)
\end{tabular}
\caption{Final binary segmentation results corresponding to (a) the amplitude image, (b) the coherence image, and (c) the fusion of amplitude and coherence of the EC dataset.}\label{fig:EC_PM_BinaryMaps}
\end{figure}

Figure \ref{fig:EC_PM_nlinsar_results} shows the coastline detection results achieved for the EC dataset using only the amplitude image, only the coherence image, and the proposed $K$-Medians-based signal-level fusion, respectively. Figure \ref{fig:EC_PM_Heatmaps} shows the corresponding maps displaying the sum of the binary maps achieved on the individual scales, which are then fed into the final decision fusion step described by equation \ref{eq:DecisionFusion}. Finally, the binary maps achieved after decision fusion and morphological filling are shown in Fig. \ref{fig:EC_PM_BinaryMaps}.

\begin{figure}[htb]
\centering
\begin{tabular}{ccc}
\includegraphics[width = 0.3\textwidth]{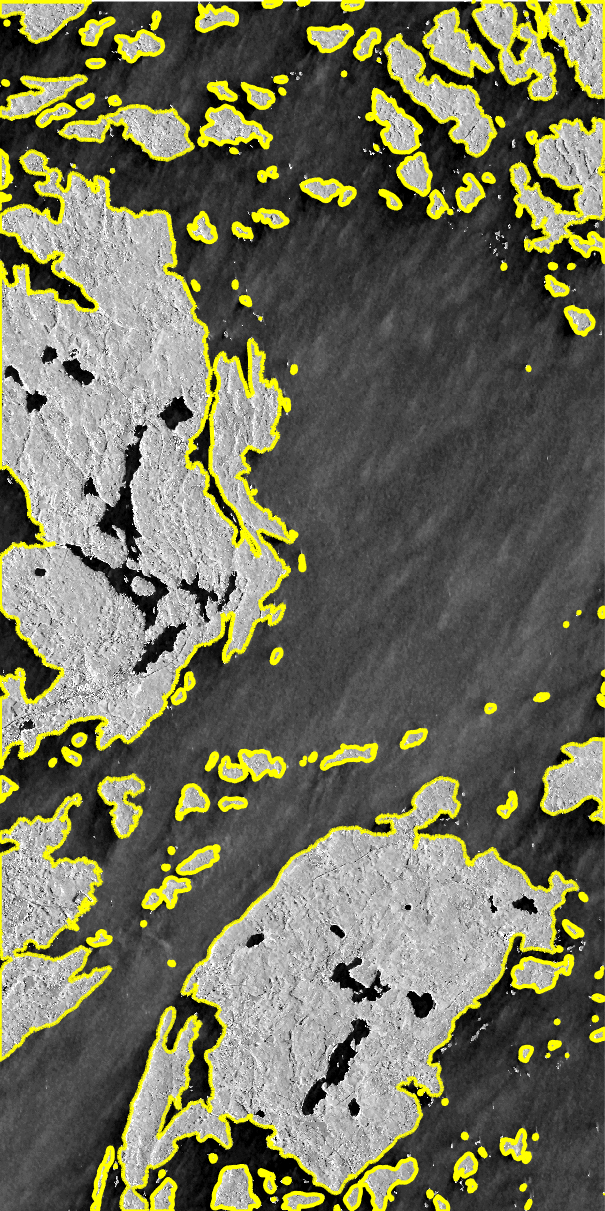} & \includegraphics[width = 0.3\textwidth]{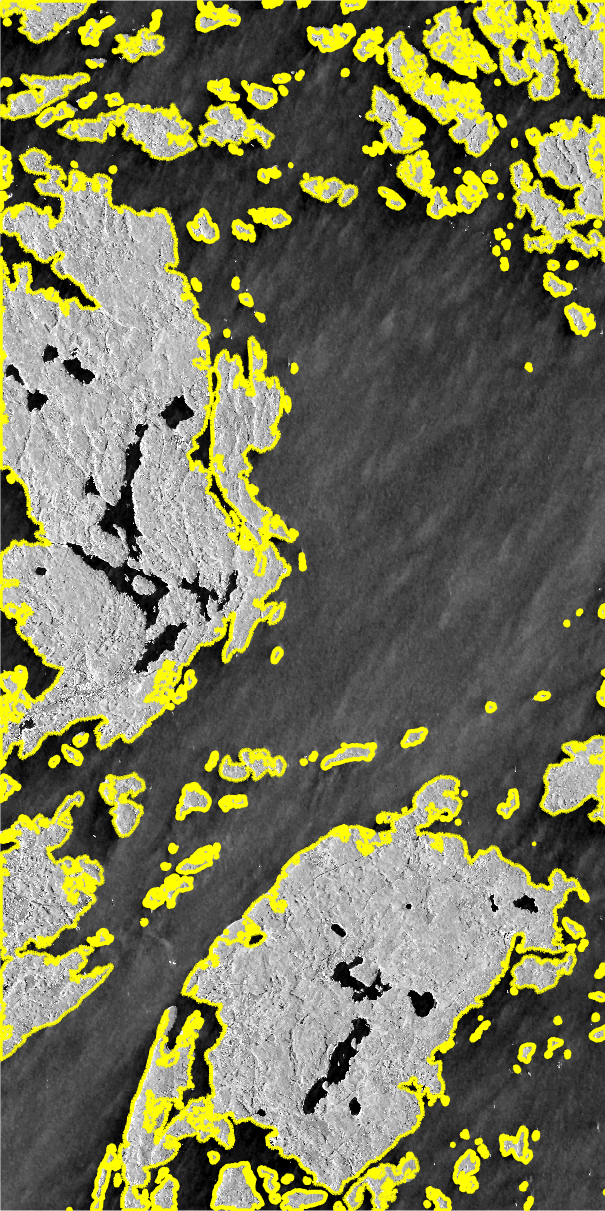} & \includegraphics[width = 0.3\textwidth]{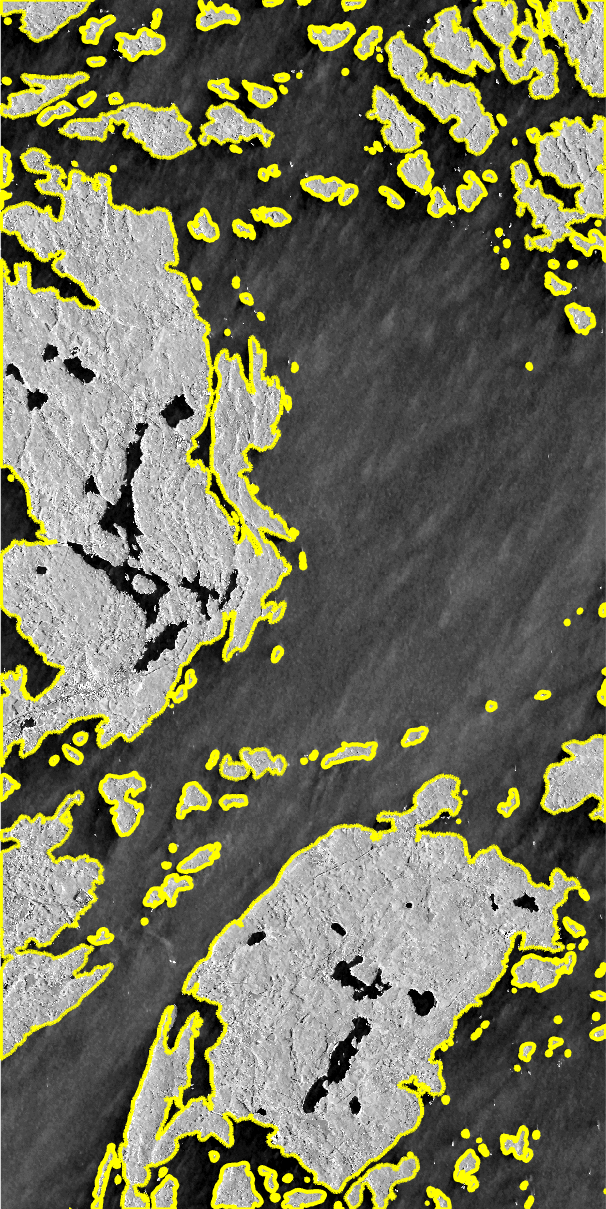}\\
(a) & (b) & (c)
\end{tabular}
\caption{Coastlines extracted from (a) the amplitude image, (b) the coherence image, and (c) from a fusion of amplitude and coherence of the ST dataset. The coastlines are overlayed to the amplitude image in all three cases.}\label{fig:ST_PM_nlinsar_results}
\end{figure}

\begin{figure}[htb]
\centering
\begin{tabular}{ccc}
\includegraphics[width = 0.3\textwidth]{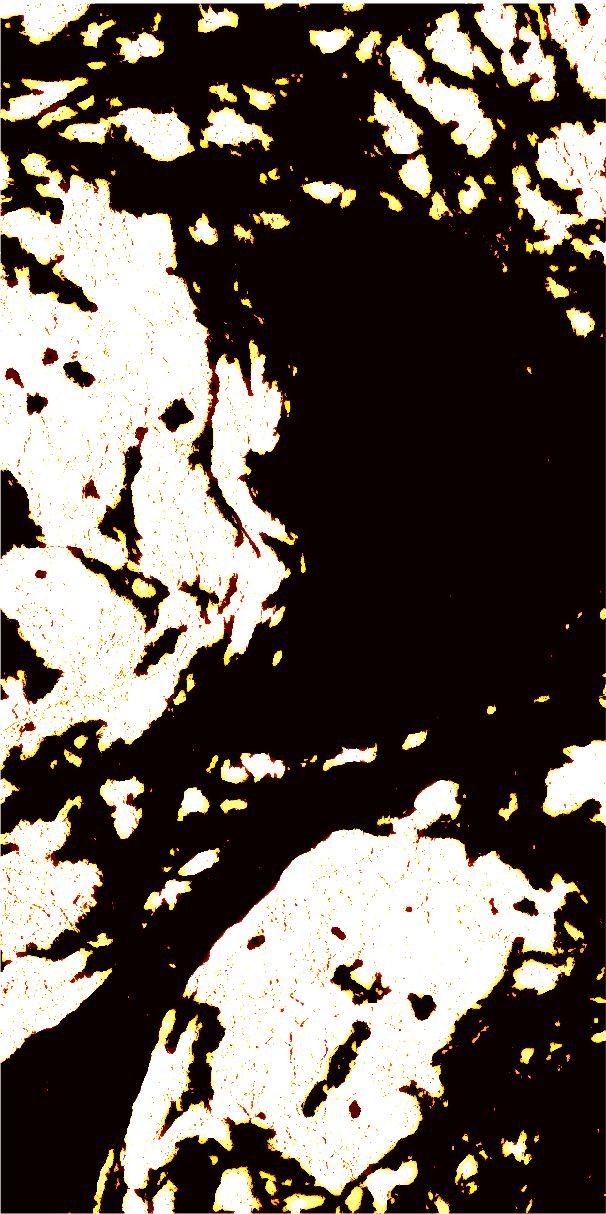} & \includegraphics[width = 0.3\textwidth]{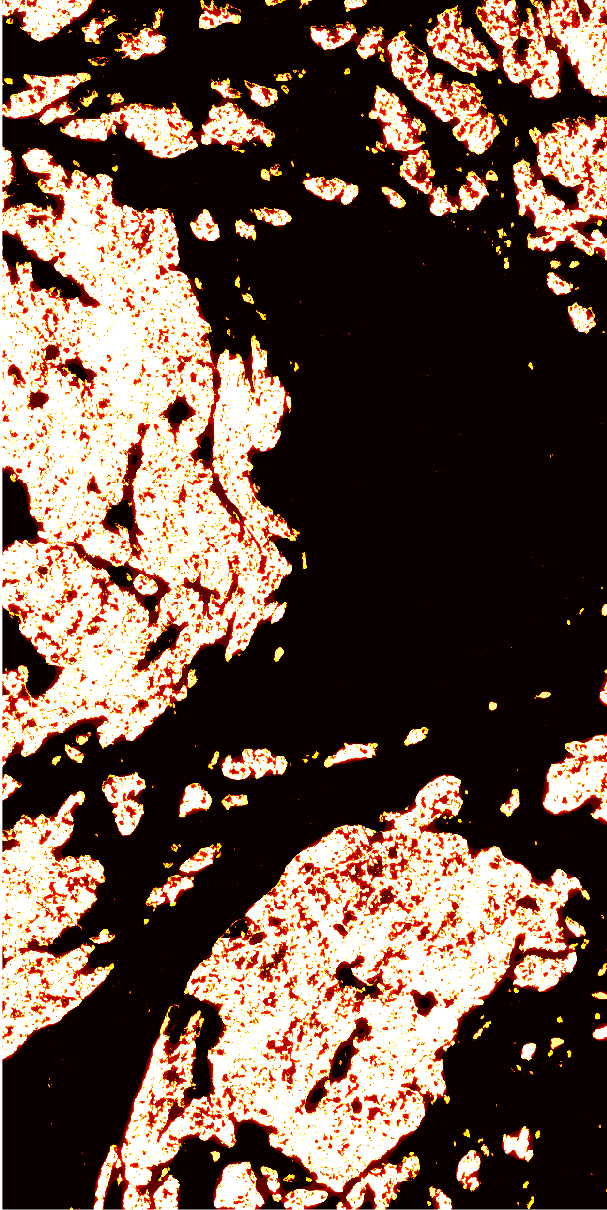} & \includegraphics[width = 0.3\textwidth]{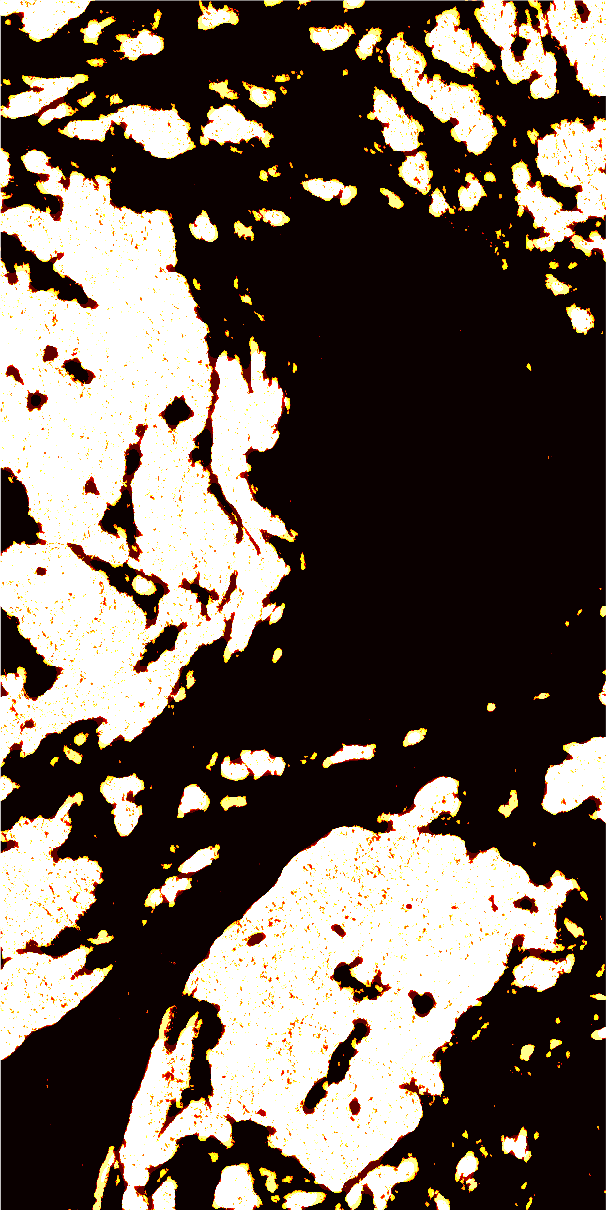}\\
(a) & (b) & (c)\\
\multicolumn{3}{c}{\includegraphics[width=0.9\textwidth]{HeatMapColorbar.png}}
\end{tabular}
\caption{Sum of the binary segmentation results corresponding to the individual scales of the scale-space representation. The maps correspond to (a) the amplitude image, (b) the coherence image, and (c) the fusion of amplitude and coherence of the ST dataset.}\label{fig:ST_PM_Heatmaps}
\end{figure}

\begin{figure}[htb]
\centering
\begin{tabular}{ccc}
\includegraphics[width = 0.3\textwidth]{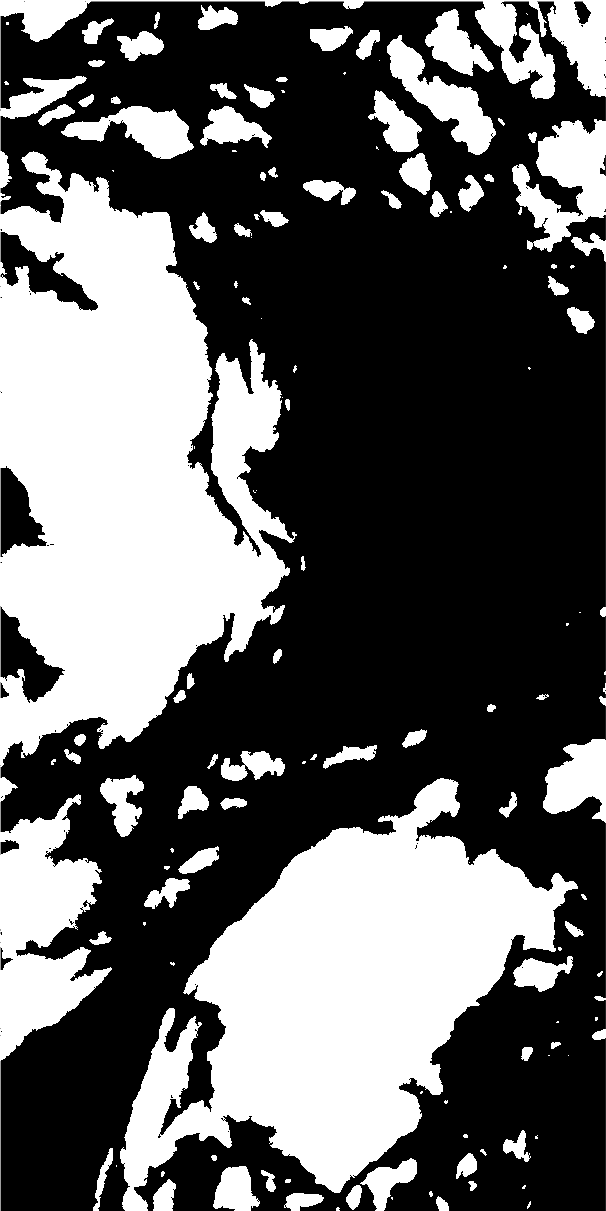} & \includegraphics[width = 0.3\textwidth]{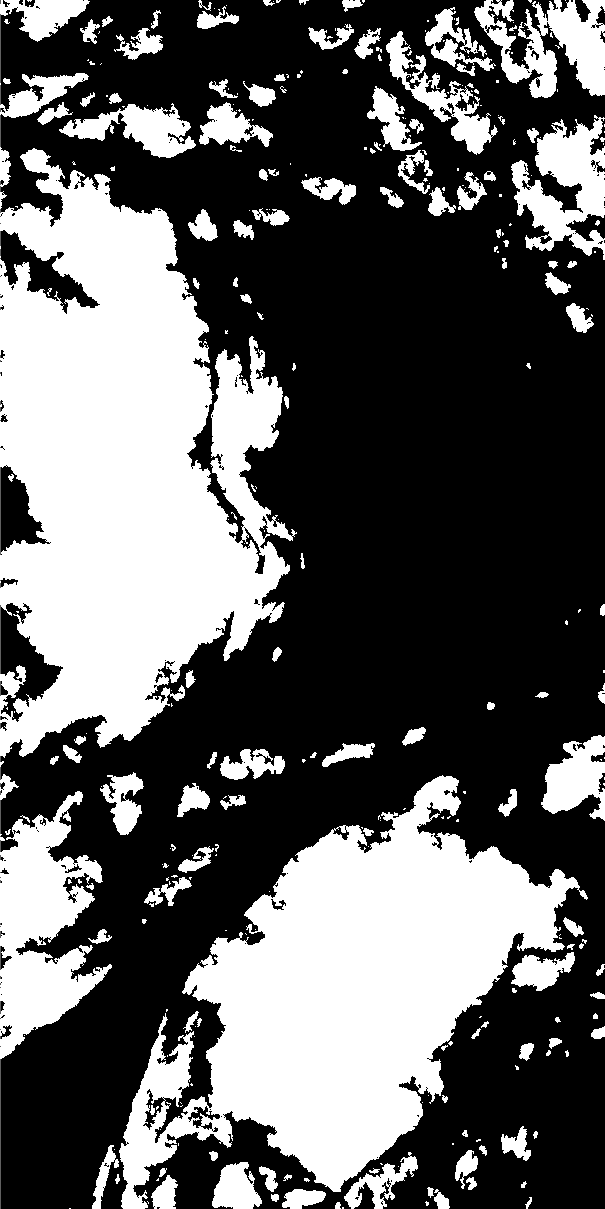} & \includegraphics[width = 0.3\textwidth]{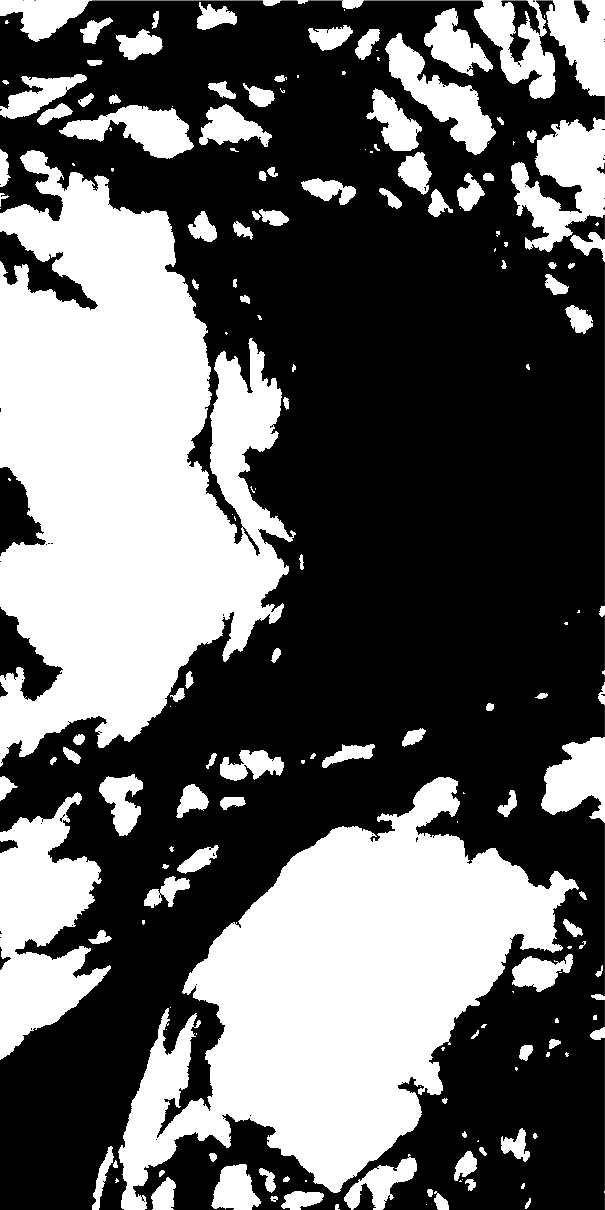}\\
(a) & (b) & (c)
\end{tabular}
\caption{Final binary segmentation results corresponding to (a) the amplitude image, (b) the coherence image, and (c) the fusion of amplitude and coherence of the ST dataset.}\label{fig:ST_PM_BinaryMaps}
\end{figure}

In a similar manner, the final coastline extraction results as well as intermediate processing results for the ST dataset are shown in Figs. \ref{fig:ST_PM_nlinsar_results}, \ref{fig:ST_PM_Heatmaps}, and \ref{fig:ST_PM_BinaryMaps}, respectively.

\begin{figure}[htb]
\centering
\begin{tabular}{ccc}
\includegraphics[width=0.3\textwidth]{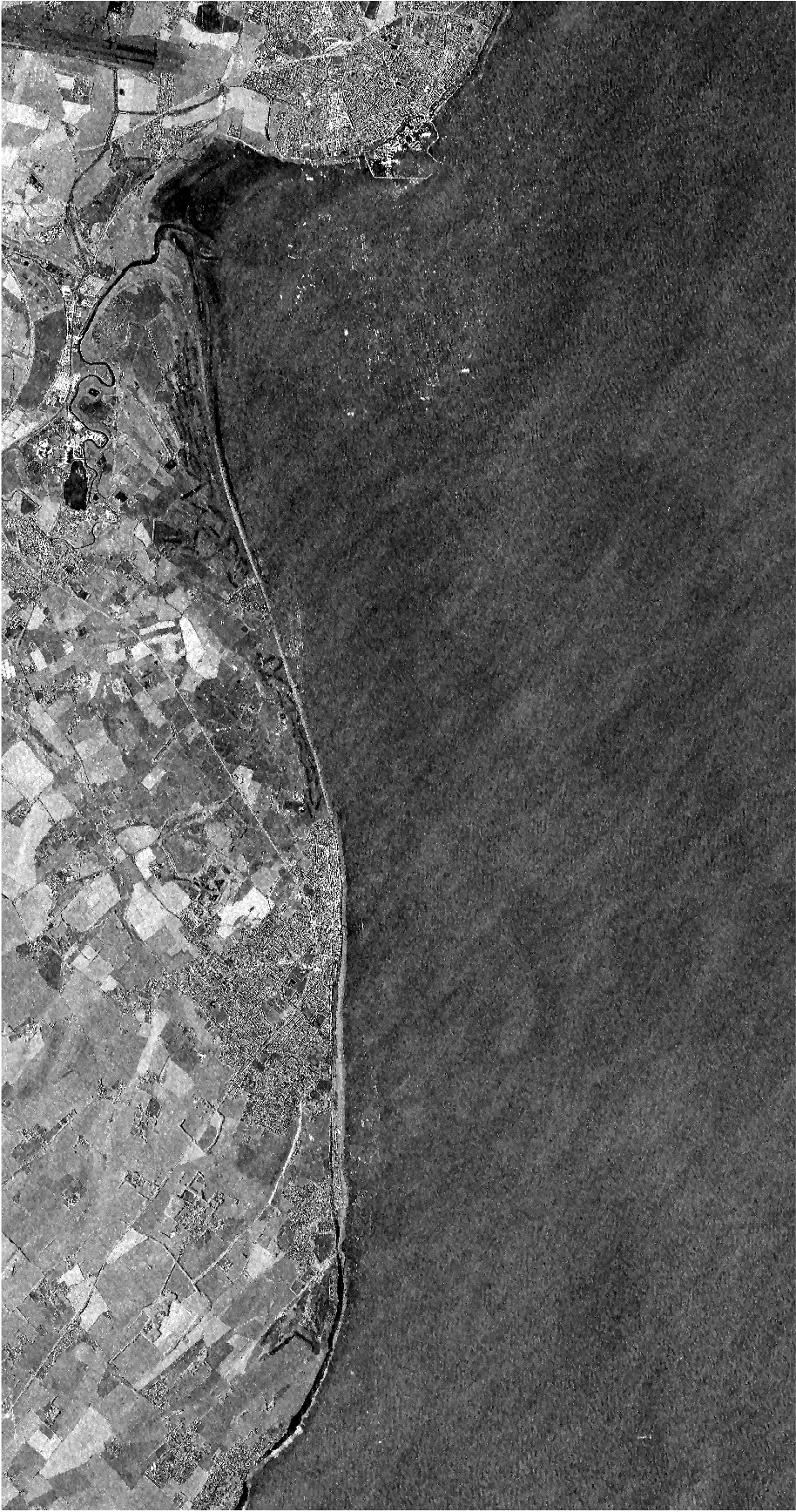} & \includegraphics[width=0.3\textwidth]{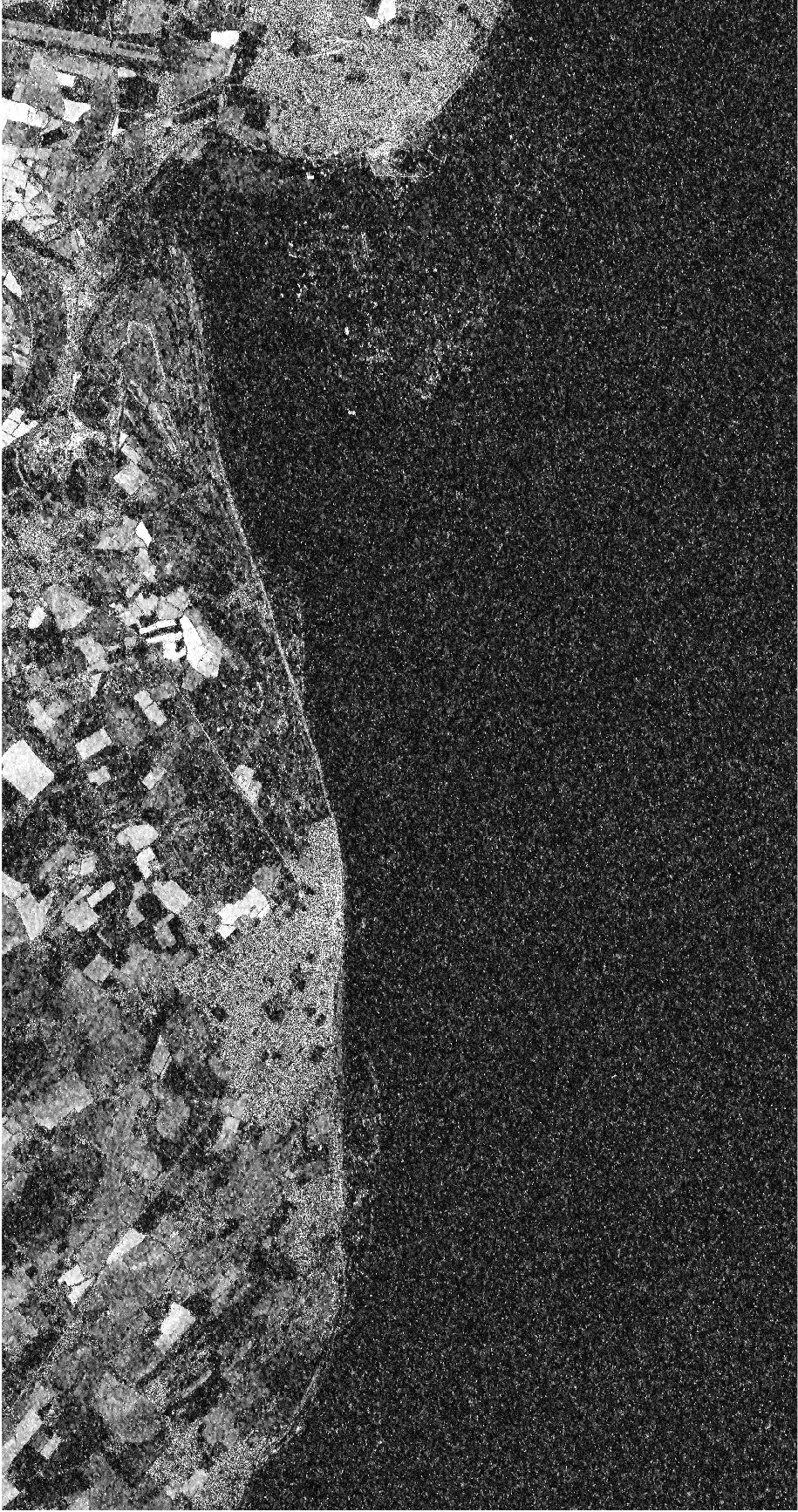} & \includegraphics[width=0.3\textwidth]{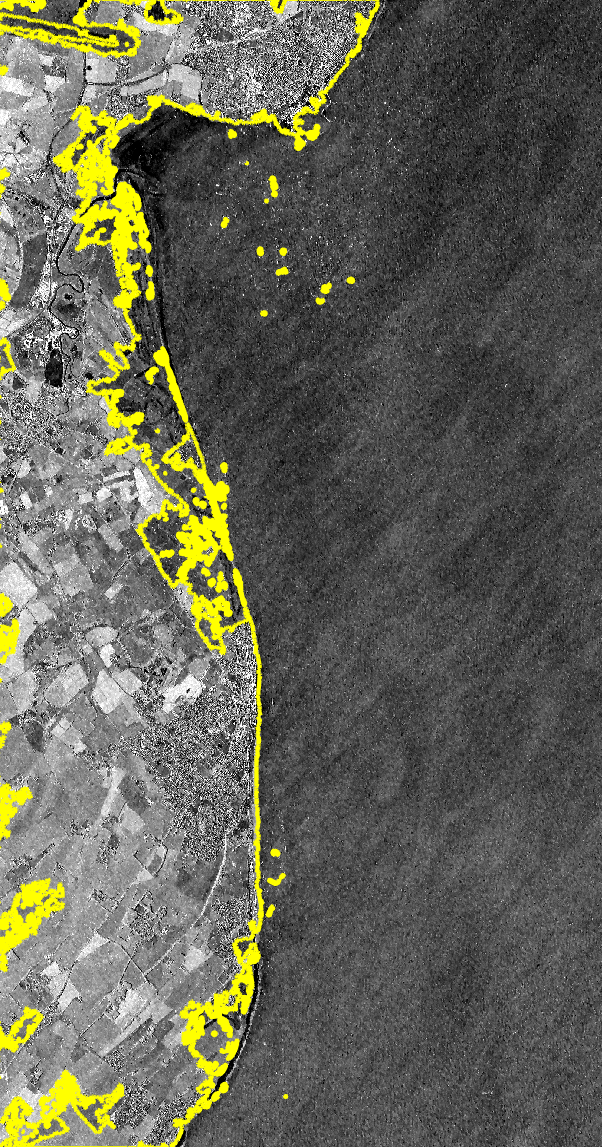}\\
(a) & (b) & (c)\\
\includegraphics[width=0.3\textwidth]{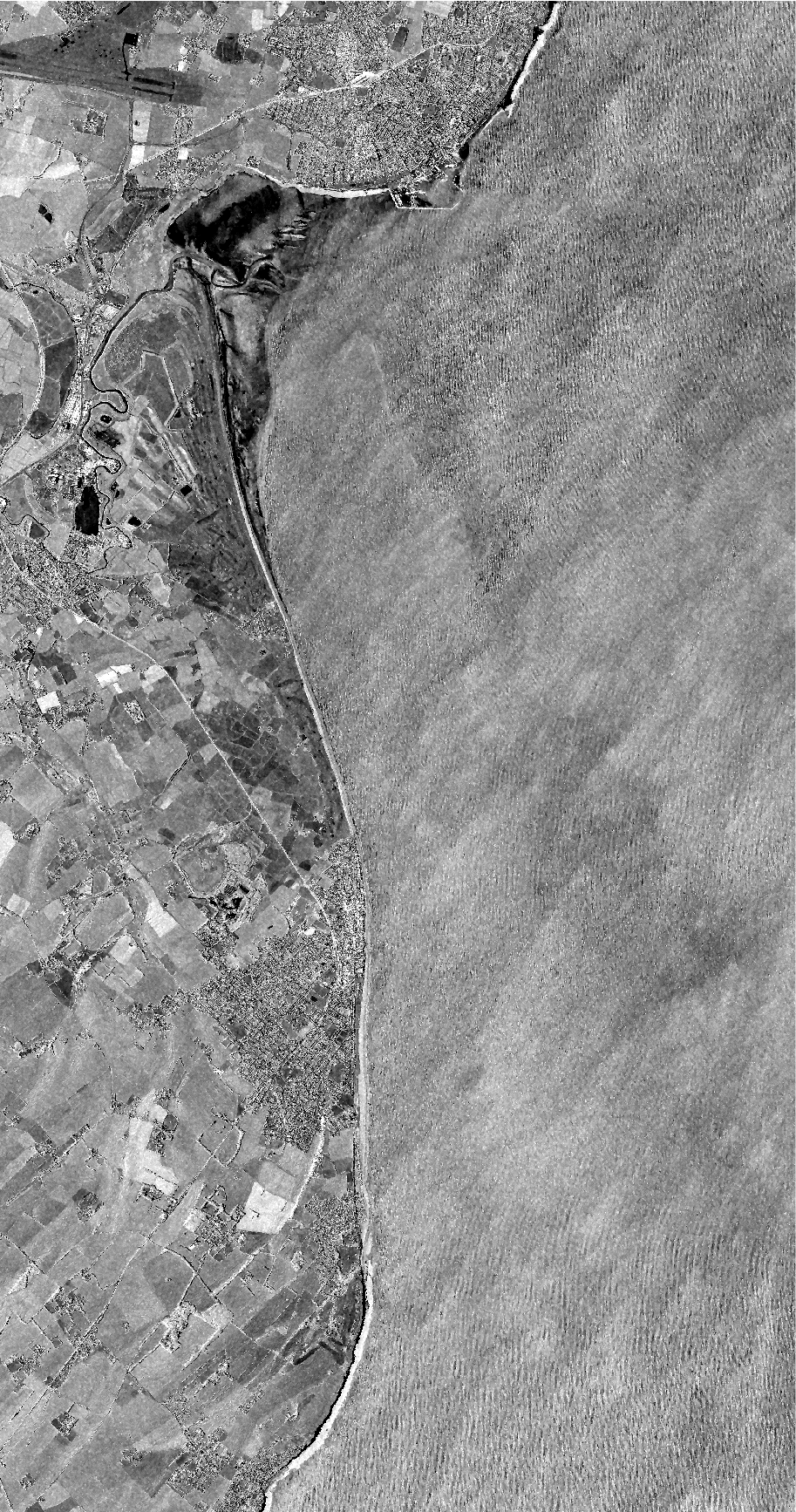} & \includegraphics[width=0.3\textwidth]{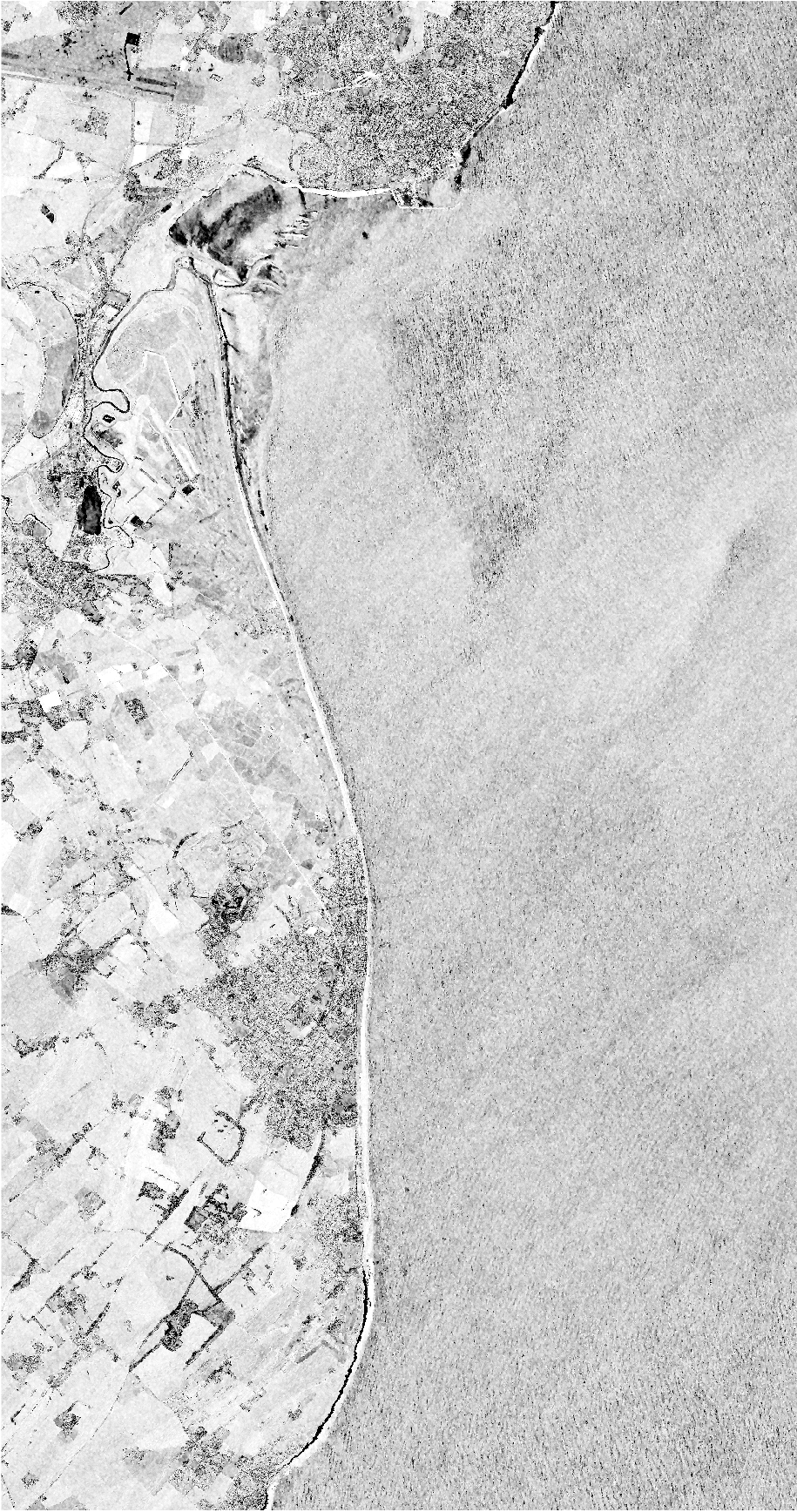} & \includegraphics[width=0.3\textwidth]{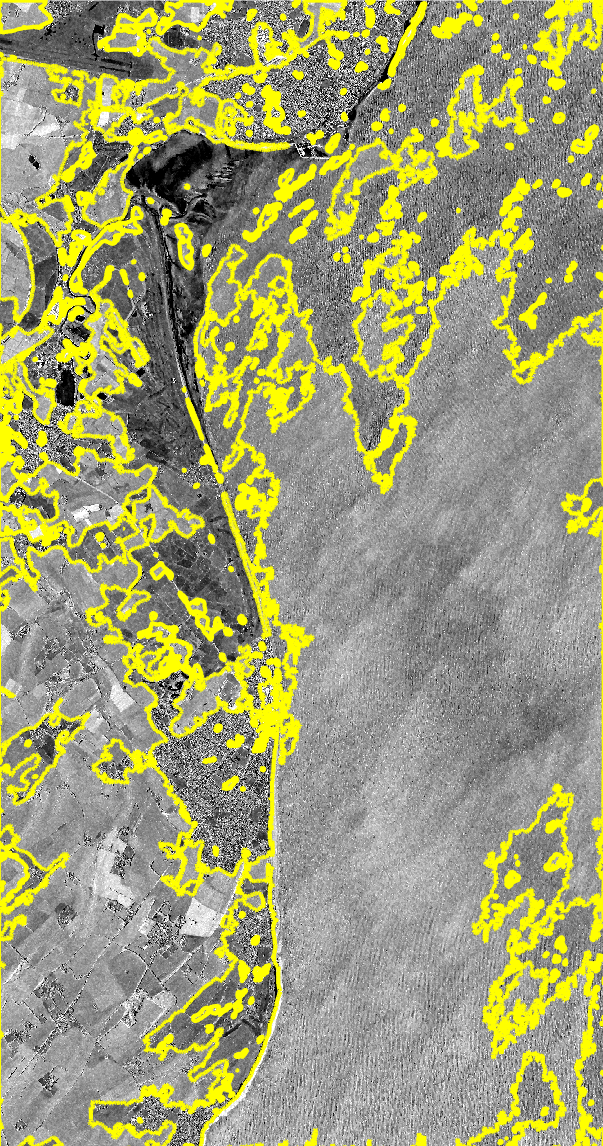}\\
(d) & (e) & (f)
\end{tabular}
\caption{Qualitative results for coastline extraction applied to data of the English Channel scene acquired in repeat-pass mode (top row) and bistatic mode (bottom row), respectively. Subfigures (a),(d) show the amplitude imagery, subfigures (b),(e) show the coherence imagery, and subfigures (c),(f) show the coastline extracted based on amplitude-coherence fusion, overlayed to the respective amplitude images.\newline Note how land surfaces appear rather dark in the repeat-pass coherence image, while rough water surfaces appear very bright in the bistatic coherence image.}\label{fig:classicResults}
\end{figure}

For sake of completeness, exemplary qualitative coastline extraction results using nonlocally filtered TanDEM-X data of the English Channel scene acquired in the classic repeat-pass and bistatic modes are displayed in Fig.~\ref{fig:classicResults}. Although we already display the more robust amplitude-coherence-fusion results, the large amount of outliers becomes clearly visible. This is caused by the effects already explained in Fig.~\ref{fig:CoherenceMatrix}: While many land surfaces decorrelate for the large temporal baseline of the repeat-pass data and thus cause confusion with water, the wind-affected water surface appears rather bright in the zero-temporal-baseline bistatic data and thus causes confusion with land. This confirms the benefit of interferometric SAR data acquired in pursuit monostatic mode for applications that aim at a simple and robust distinction between water and land surfaces.

\begin{figure}[htb]
\centering
\begin{tabular}{cc}
\includegraphics[height = 10cm]{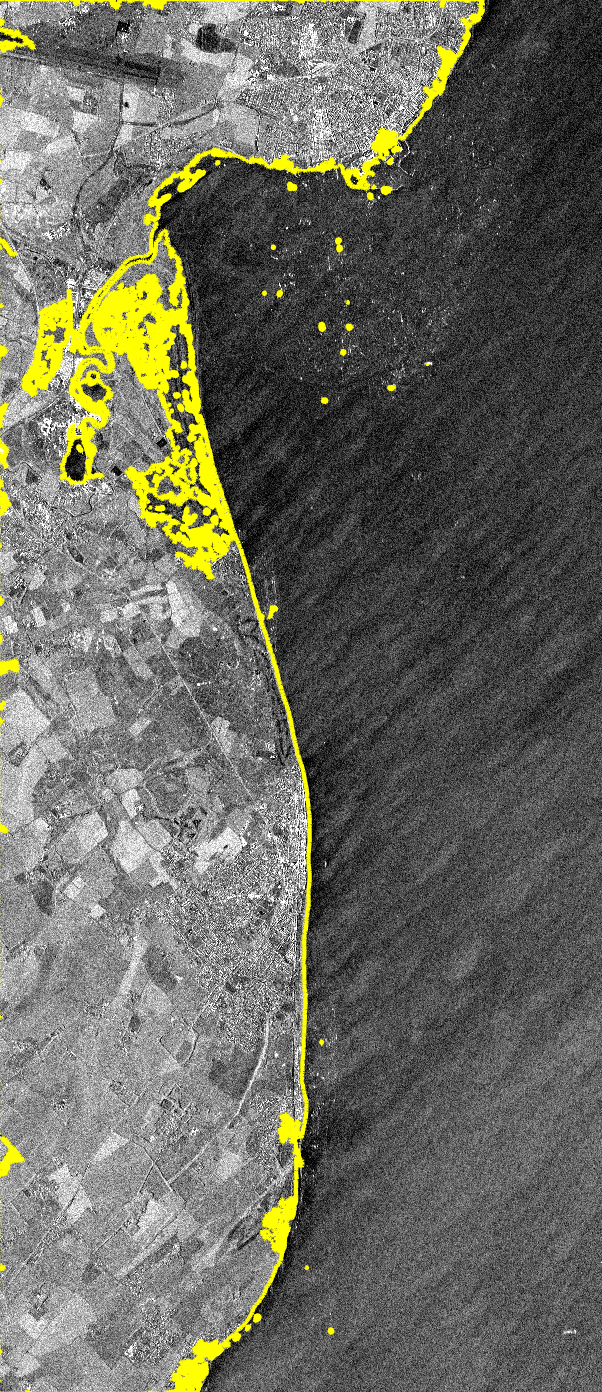} &  \includegraphics[height = 10cm]{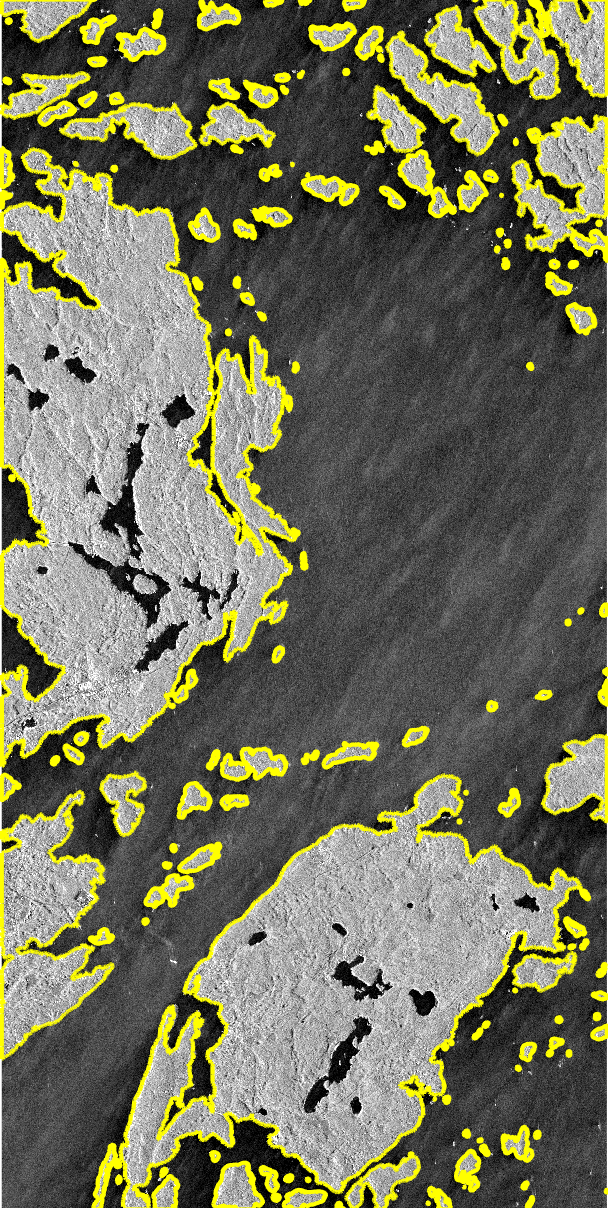}\\
(a) & (b) 
\end{tabular}
\caption{Coastlines extracted from a fusion of amplitude and coherence of a boxcar-filtered version of (a) the EC dataset and (b) the ST dataset. The coastlines are overlayed to the amplitude image in both cases.}\label{fig:boxcar_results}
\end{figure}

Last, but not least, Fig.~\ref{fig:boxcar_results} displays the boxcar-filter-based results of coastline detection on the fused amplitude and coherence imagery acquired in pursuit monostatic mode for both the EC and the ST cases. It becomes immediately apparent that many outliers exist, especially on the land surface of the EC dataset. This is caused by the lower filtering power of the boxcar filter along the coastline. 

\subsection{Quantitative Results}
From visual inspection of the coastline extraction results presented in Sections~\ref{sec:qualResulPM} it becomes apparent that the fusion of nonlocally filtered amplitude and coherence imagery acquired in pursuit monostatic mode is able to deliver the most reliable results for both test scenes. However, we additionally carry out a quantitative evaluation with respect to a manually generated reference dataset for both nonlocally filtered and boxcar filtered data to further investigate the benefit of nonlocal filtering. The results are displayed in Tab.~\ref{tab:Results}.

\begin{table}[htb]
\centering
 \footnotesize
  \caption{Statistics of the euclidian distances (in m) between the detected coastline and the reference coastline}\label{tab:Results}
\begin{tabular}{lccc}
\toprule
Experiment & $25\%$-quantile & $50\%$-quantile & $75\%$-quantile\\
~~\textbf{English Channel} &&&\\
~~~~\textit{NLInSAR} & 3.2 & 9.0& 26.2\\
~~~~\textit{Boxcar} & 4.0 & 16.2 & 34.0\\
~~\textbf{Stockholm} &&&\\
~~~~\textit{NLInSAR} & 1.0 & 2.2 & 6.4\\
~~~~\textit{Boxcar} & 1.0 & 3.0 & 10.0\\
\bottomrule
\end{tabular}
\end{table}

In this context, we have to comment on the common coastline evaluation approaches: As mentioned in Section~\ref{sec:CoastlineDetection}, quite a number of different coastline detection algorithms have been published since 1990. Given the high costs associated with ground truth generation in difficult terrain as well as the difficulty to match an extracted coastline with a reference coastline, it is interesting to have a look at the different ways that have been employed for evaluation of the individual coastline detection results. Immediately, it comes to attention that a significant share of all papers on SAR-based coastline detection do not provide any quantitative evaluation and only evaluate the resulting coastline by (visually) comparing it to the SAR image that was processed \cite{Lee1990,Zhang1994,Descombes1996,Mason1996,Horritt1999,Niedermeier2000,Niedermeier2005,Silveira2009,Baselice2013,Nunziata2014,Wiehle2015}. Only few authors go one step further and compare their results with external reference data, e.g. derived from a topography model and the average tide \cite{Schwaebisch1997}, a reference coastline extracted from optical imagery or topographic maps \cite{Dellepiane2004,Liu2004}, or from sparsely distributed GPS control points \cite{Nunziata2016,Liu2017}. 

Besides generating an objective reference, the second major reason for only marginal evaluations so far lies in the difficulty to properly match an extracted coastline to a reference coastline, especially if the coast under investigation is not relatively smooth and quasi-linear, but follows a more complicated pattern. 
Thus, we followed the approach of Dellepiane et al. \cite{Dellepiane2004}, who proposed to convert the reference coastline into a distance image, from which subsequently accuracy statistics can be extracted. We combine this approach with a manual selection of the coastline parts to be evaluated in order to remove a potential quantitative bias caused by completely qualitatively wrong detections. The distances, which are originally retrieved in pixels, are then multiplied by the pixel size of the georeferenced image in order to get the metric distance between extracted and reference coastlines. Finally, we calculate the median values from the retrieved distances to further robustify the assessment.

\section{Discussion}\label{sec:discussion}
\begin{figure}[htb]
\centering
\begin{tabular}{cc}
\includegraphics[width=0.3\textwidth]{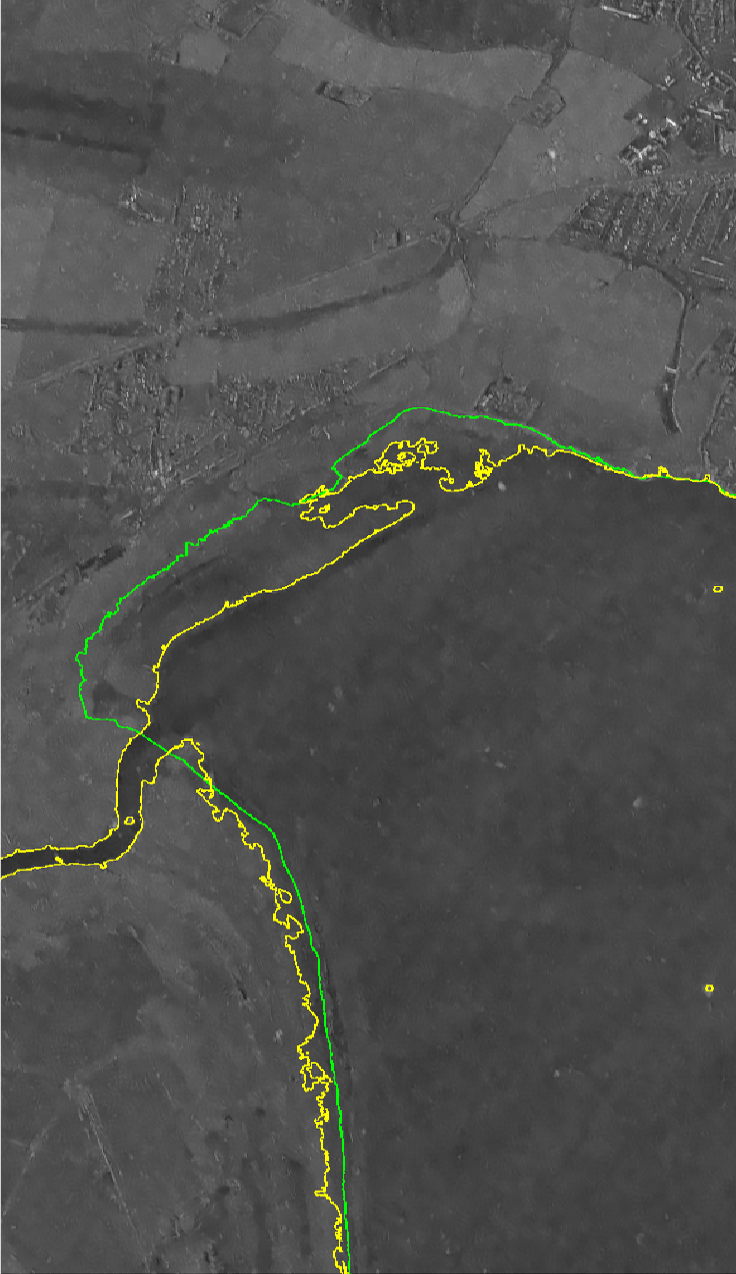} & \includegraphics[width=0.3\textwidth]{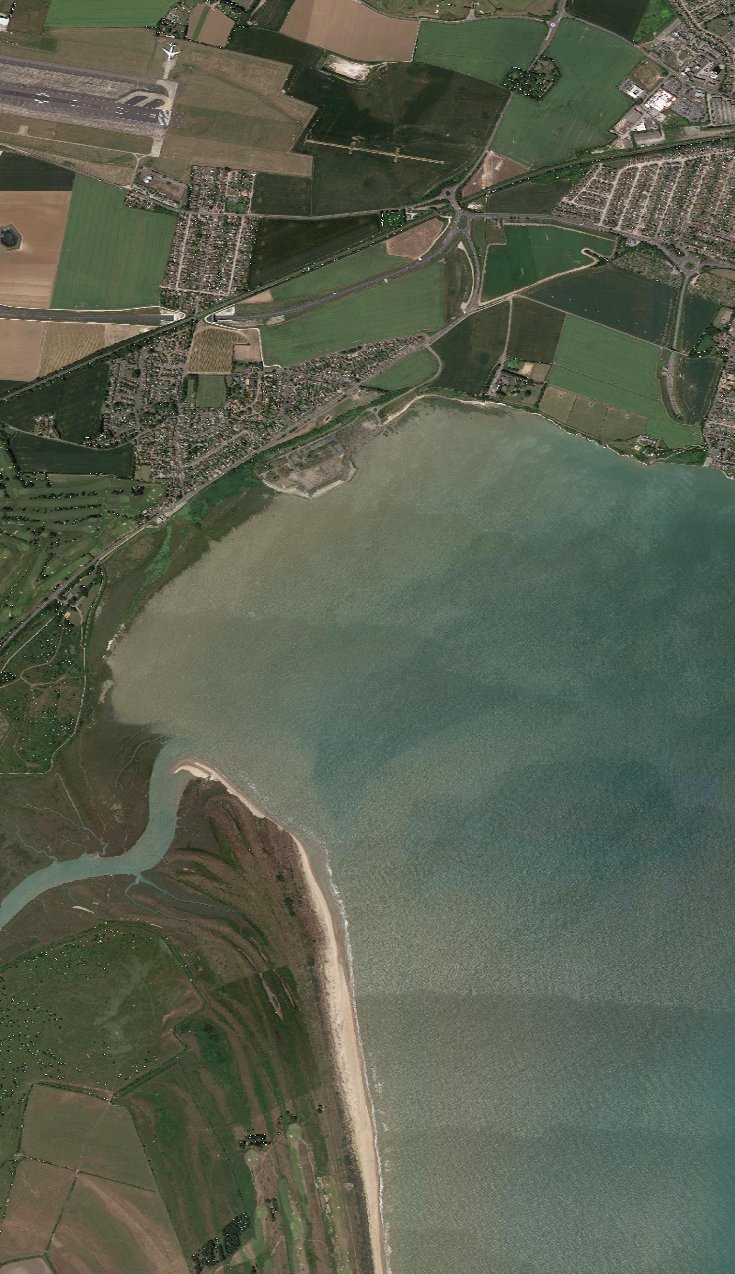}\\
(a) & (b)\\
\includegraphics[width=0.3\textwidth]{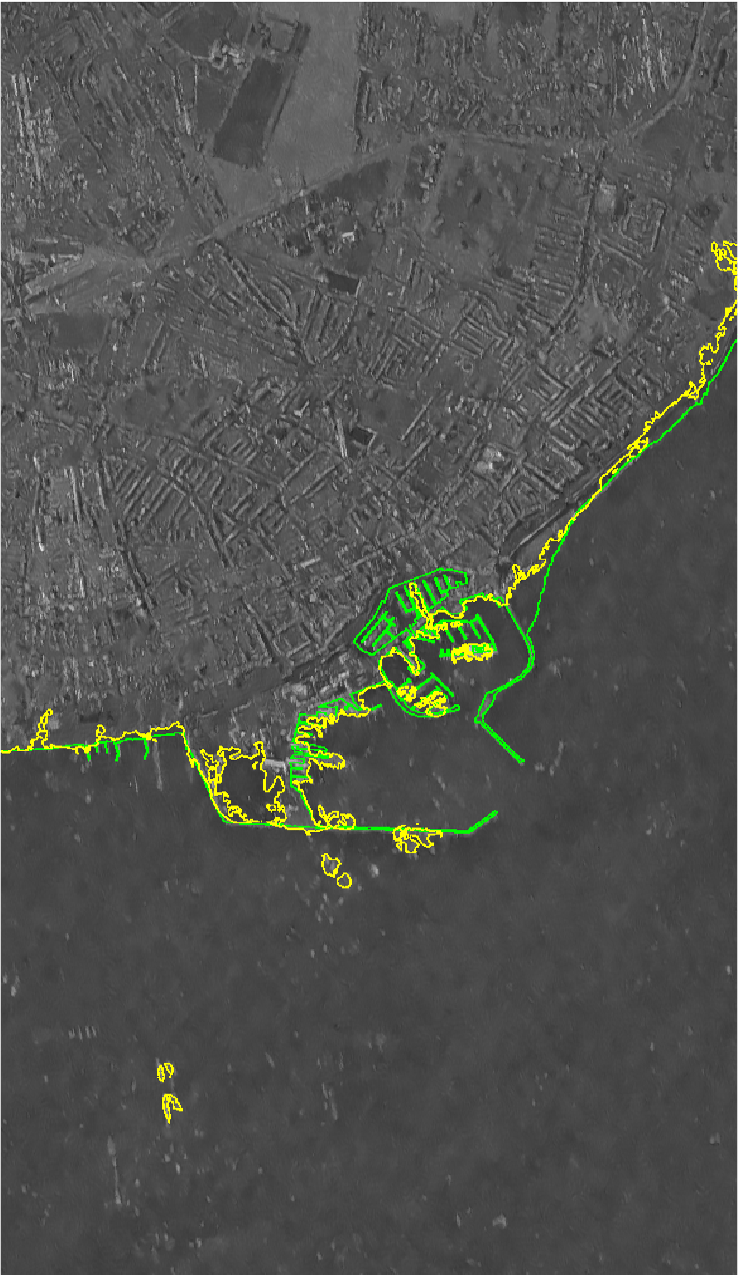} & \includegraphics[width=0.3\textwidth]{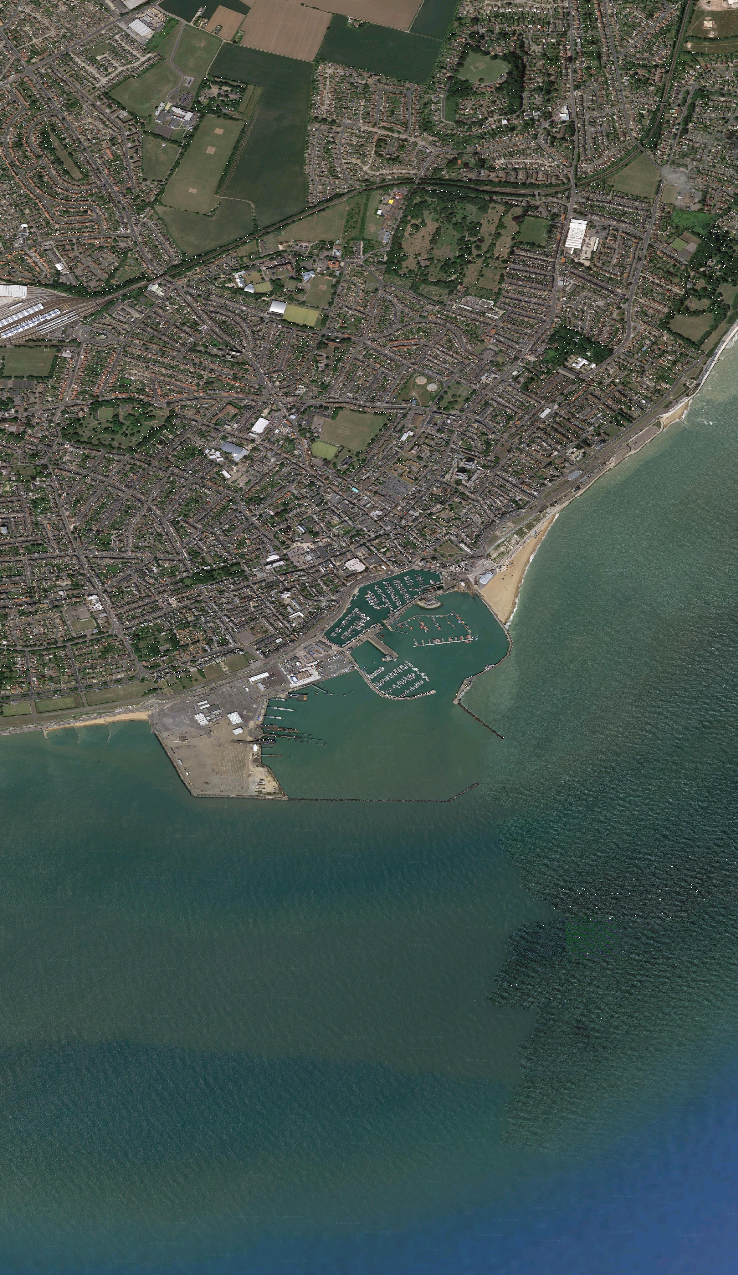}\\
(c) & (d)\\
\includegraphics[width=0.3\textwidth]{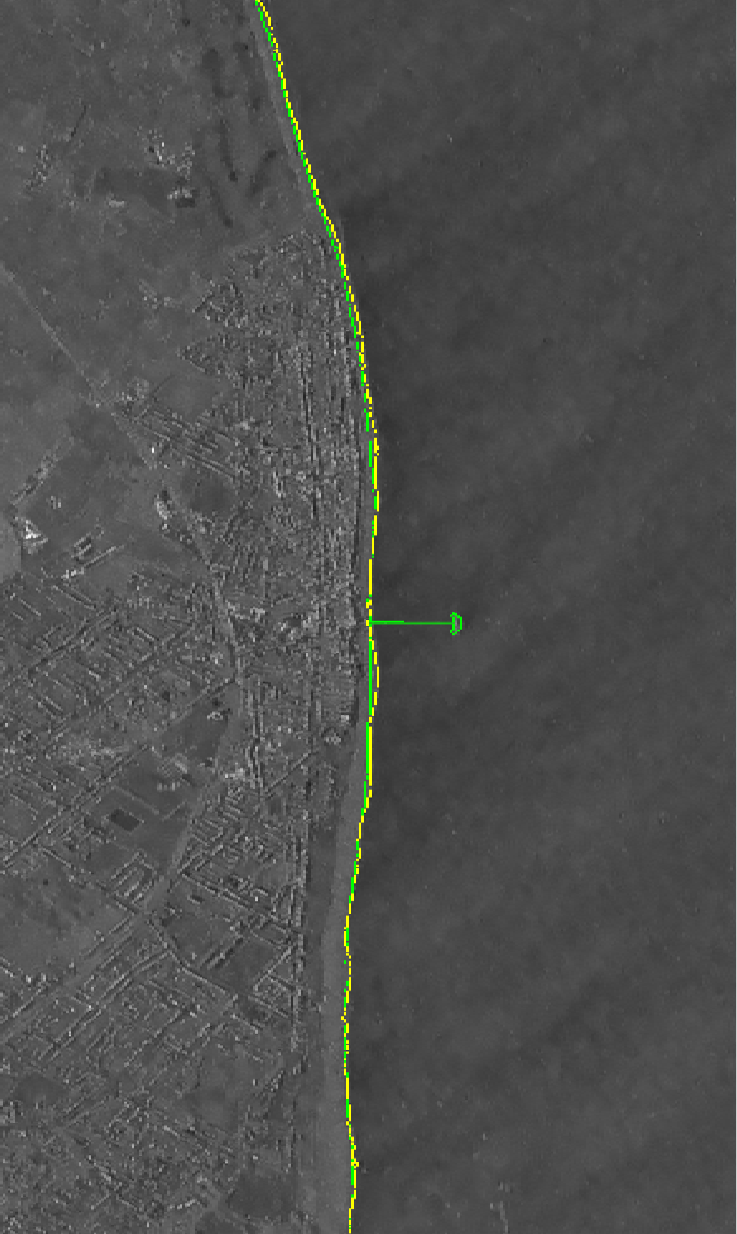} & \includegraphics[width=0.3\textwidth]{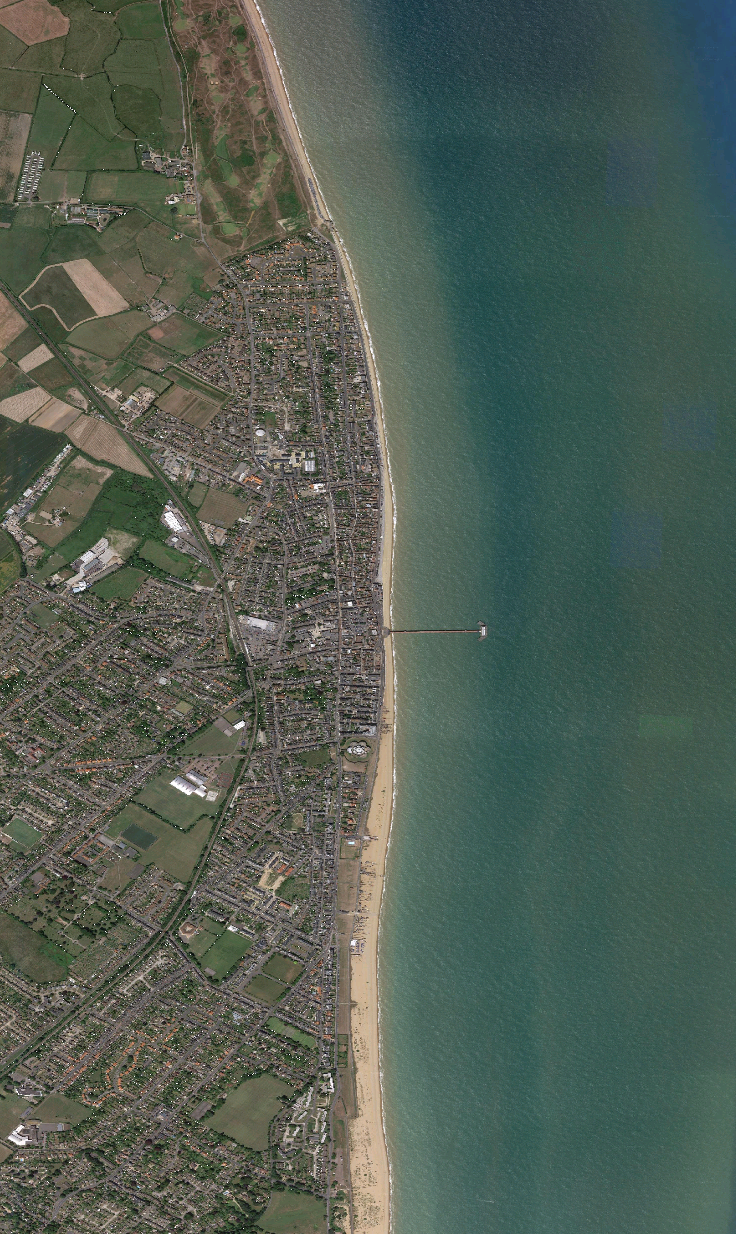}\\
(e) & (f)
\end{tabular}
\caption{Exemplary subsets of the EC dataset: Comparison of optical satellite images with the reference coastlines extracted therefrom (overlayed in green onto the corresponding SAR amplitude image) and the coastline extracted by our approach (overlayed in yellow). This shows some exemplary deviations, which are caused by shallow water areas close to the beach (a,b), by detailed harbour piers existing in the reference data, but not detected in such detail by the automatic approach (c,d), and by a jetty, which also was not detected by our coastline extraction approach (e,f).}\label{fig:EC_eval}
\end{figure}

\begin{figure}[htb]
\centering
\begin{tabular}{cc}
\includegraphics[width=0.3\textwidth]{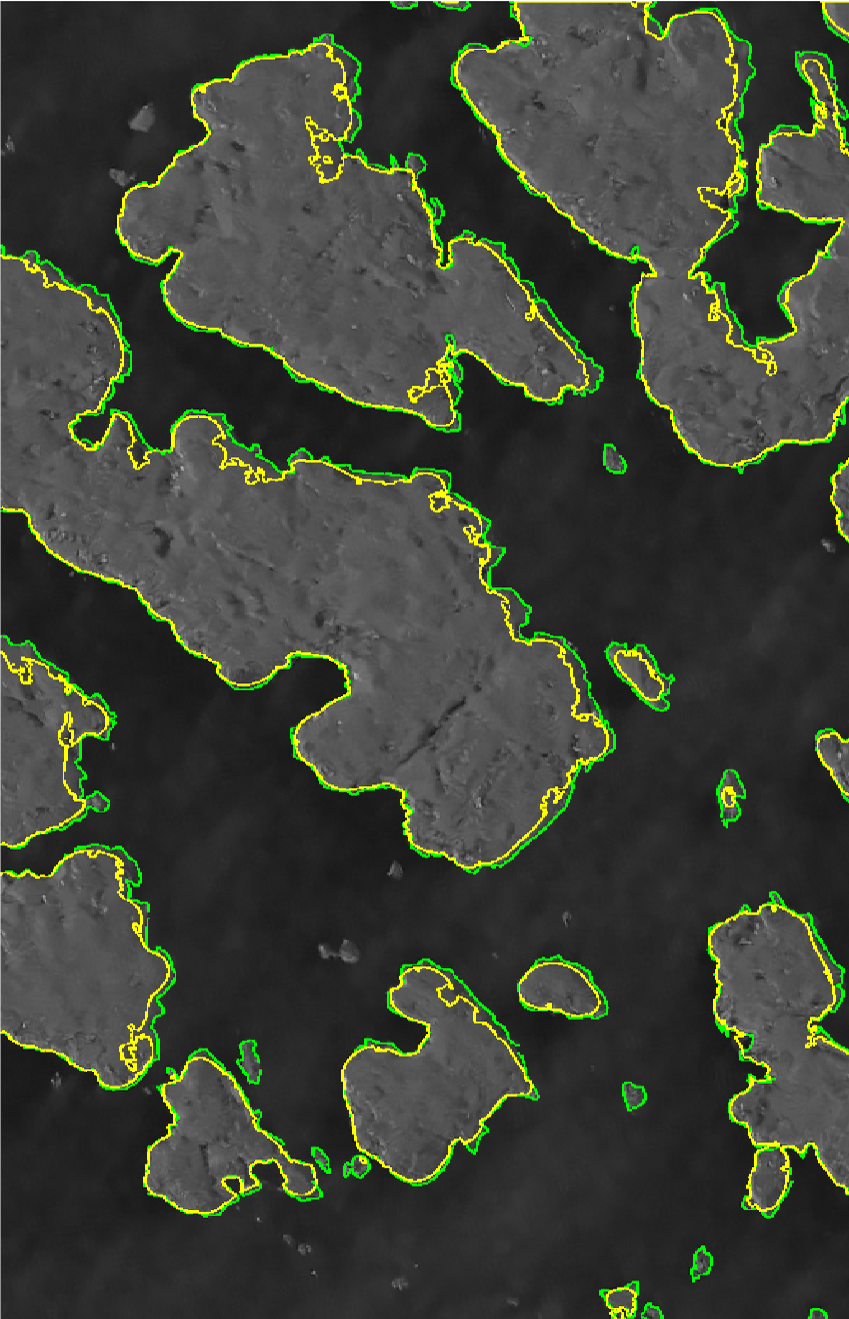} & \includegraphics[width=0.3\textwidth]{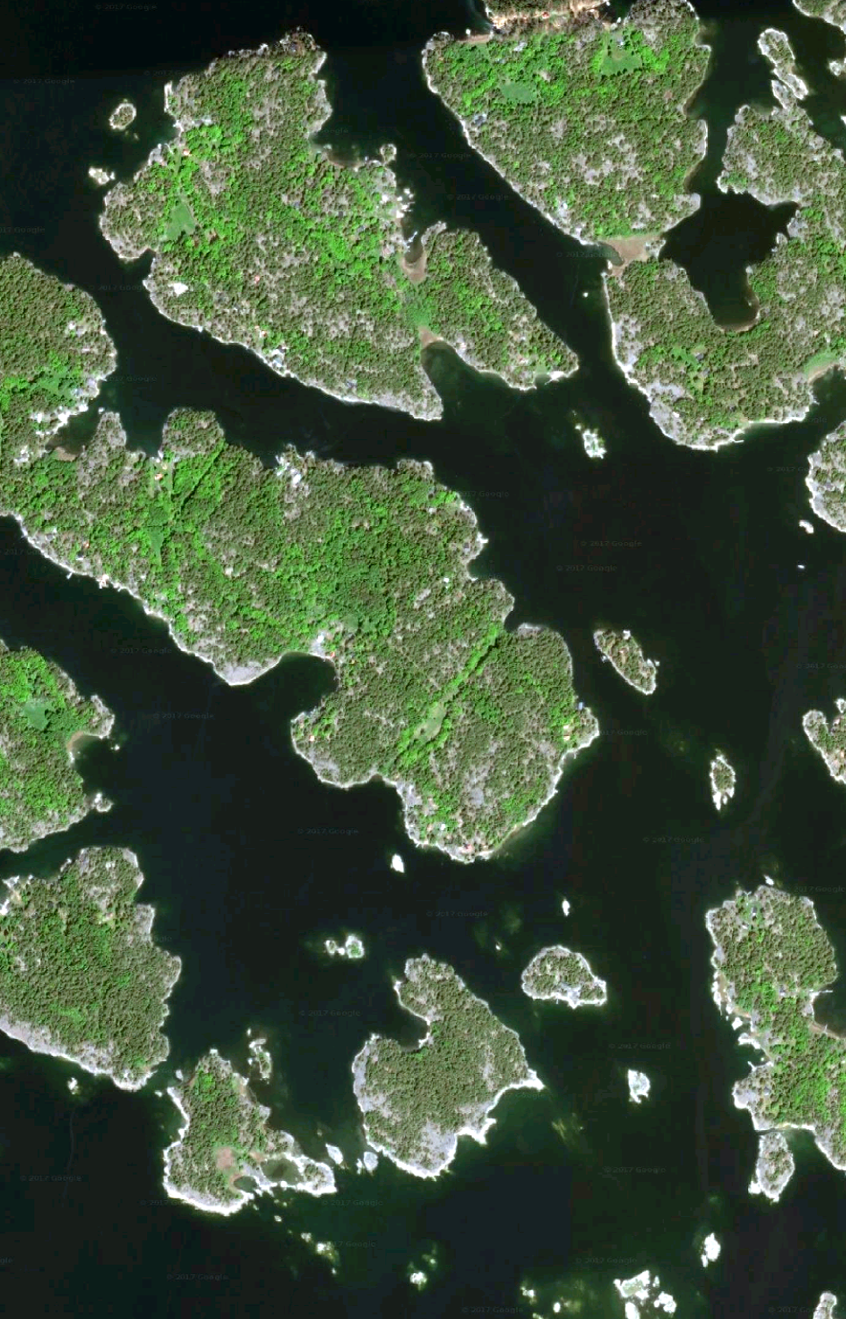}\\
(a) & (b)\\
\end{tabular}
\caption{Exemplary subset of the ST dataset: Comparison of optical satellite images with the reference coastlines extracted therefrom (overlayed in green onto the corresponding SAR amplitude image) and the coastline extracted by our approach (overlayed in yellow). This comparison exemplifies the good accordance between reference coastline and extracted coastline. While some very small skerries are not extracted, they are also not contained in the reference dataset. Besides few missed smaller islands, the remaining extraction result follows the reference coastline in good detail.}\label{fig:ST_eval}
\end{figure}

The experiments presented in Section \ref{sec:experiments} prove that interferometric TanDEM-X data provide a rich source of information for coastline extraction purposes especially if the following conditions are met:
\begin{itemize}
\item[1)] The data are acquired in pursuit-monostatic mode. Both theoretical considerations (cf. Section~\ref{sec:TDXpm}) as well as our study results indicate that the decorrelation of water surfaces and the comparatively high temporal coherence of land surfaces leads to improved separability between the two classes.
\item[2)] Nonlocal filtering enhances the achievable coastline extraction accuracies with respect to conventional boxcar filtering. The benefit mainly arises in inhomogeneous areas close to the actual coastline, where otherwise land-water segmentation and the subsequent morphological filling process fail and thus eventually lead to coastline misdetections.
\item[3)] Although both amplitude and coherence images already provide a good information source for coastline extraction on their own, fusing them on a signal-level helps to improve land-water-segmentation and thus coastline detection even further. 
\end{itemize}
Using a simple, fully automatic coastline extraction method, which exploits both scale space representations and $K$-medians from an algorithmic point-of-view and signal-level fusion as well as decision-level fusion from a data fusion point-of-view, we achieved reliable coastline extraction results that show median distances to the reference coastlines, which were manually extracted from optical remote sensing data, between 2.2m and 9.0m with nonlocal filtering and 3.0m and 16.2m with boxcar filtering. Comparing these results to the few pieces of related work, which provide a quantitative evaluation, this seems to be quite promising: Liu et al. \cite{Liu2004} report a mean distance of 28 m for sample points extracted from topographic maps and 105 m for a reference coastline extracted from optical SPOT imagery, while 
Nunziata et al. \cite{Nunziata2016} achieved an accuracy between 25m and 100m with respect to GPS measurements acquired for a small subset of the study area. Most recently, Liu et al. \cite{Liu2017} calculated a mean offset of 10m to 15m between their extracted coastlines and a set of manually traced ground points whose coordinates were also measured by GPS. While our EC results correspond to the state-of-the-art accuracy spectrum, the ST results are improving this state-of-the-art by about one order of magnitude. Explanations for these results can be drawn from Figs.~\ref{fig:EC_eval} and \ref{fig:ST_eval}, respectively: Although the ST dataset might seem more challenging at first glance due to the numerous islands of different sizes and the structure-rich coastlines, it proved to be less challenging for our coastline extraction approach, which benefits greatly from the adaptiveness of both the non-local filtering of the data as well as the scale-space-based segmentation approach. The EC dataset, in contrast, contains some more challenging features. As an example, Fig.~\ref{fig:EC_eval}~(a-b) shows a beach area with very shallow water, which was misclassified as land surface by our approach. The distance between the actual coastline and the extracted coastline is about 250 m. The situation is similar for the jetty displayed in Fig.~\ref{fig:EC_eval}~(e-f), which is also more than 250 m long. In contrast, the rest of the rather simple coastline fits almost perfectly. Finally, Fig.\ref{fig:EC_eval}~(c-d) shows another example that led to a deterioration of the quantitative results, as also this harbour area, whose piers were contained in the reference dataset in great detail, were not removed as outliers during the evaluation. 

This assessment shows several things: For both the EC and the ST datasets, qualitative evaluation shows very promising results, which is also confirmed by the quantitative evaluation of the ST dataset. While the quantitative results achieved for the EC dataset are still in correspondence with the state-of-the-art, they are significantly worse than the ST results. On the one hand, this deterioration can be ascribed to mis-segmentations caused by a confusion of shallow water areas and land surfaces, which indicates a data-inherent shortage that has to be considered when aiming at operational coastline extraction procedures. On the other hand, however, it can also be ascribed to significant differences between reference coastline and extracted coastline in areas containing complicated man-made structures. Thus, the achievable coastline extraction accuracy that can be expected from a combination of the approach presented in this paper and nonlocally filtered pursuit monostatic TanDEM-X data can be assumed to lie in the meter-domain.

\section{Summary \& Conclusion}\label{sec:conclusion}
The major contributions of this article are two-fold: On the one hand, we presented a simple, fully automatic and unsupervised coastline extraction procedure based on scale-space theory, $K$-medians filtering and data fusion. On the other hand, we investigated the advantage of utilizing interferometric SAR data acquired in the innovative pursuit-monostatic mode provided during the science phase of the TanDEM-X mission (rather than conventional bistatic or repeat-pass data) as promising source of information for land-water segmentation and coastline extraction tasks. In addition, we showed the benefit of nonlocal SAR data filtering over conventional boxcar filters. In essence, we were able to extract coastlines with accuracies in the meter-domain, which provides a significant improvement over the results we have achieved with conventional repeat-pass or bistatic data, and with conventional box-car filters, respectively. Thus, with respect to future mission designs, the implementation of a pursuit-monostatic mode can be recommended if applications such as automatic coastline detection are envisioned.

\section*{Acknowledgment}
The authors would like to thank the colleagues involved in the ordering, stacking, and geocoding of the TanDEM-X data: Sind Montazeri, Nan Ge, Fernando Rodriguez, and Dr. Yuanyuan Wang. Furthermore, they want to thank Rong Huang for extracting the reference coastlines from Google Earth imagery.\\
The work was funded by by the Federal Ministry for  
Economic Affairs and Energy of Germany in the project ``J Lo - The Joy of Long Baselines'' (FKZ 50EE1417).

\bibliographystyle{model1-num-names}
\bibliography{coastlines}

\end{document}